\def\rn{\noindent\parshape 2 0truecm 8.8truecm 0.3truecm 8.5truecm}
\def\nn#1 #2{#1, #2.}				
\def\nnn#1 #2 #3{#1, #2. #3.}			
\def\nnnn#1 #2 #3 #4{#1, #2. #3. #4.}		
\def\nnnnn#1 #2 #3 #4 #5{#1, #2. #3. #4. #5.}	
\def\dualand{, \&\hbox{ }}				
\def\multiand{, \&\hbox{ }}				

\def\rg#1;#2;#3;#4;#5;#6 {\par\rn#1 #2, {\it #3}, {\bf #4}, #5 (``#6'') \par}
\def\rf#1;#2;#3;#4;#5 {\par\rn#1 #2, {\it #3}, {\bf #4}, #5\par}
\def\rfbook#1;#2;#3;#4;#5 {{\frenchspacing\par\rn#1 #2, {\it #3} (#4: #5)\par}}
\def\rfproc#1;#2;#3;#4;#5;#6 {{\frenchspacing\par\rn#1 #2, in {\it #3}, ed. #4 (#5: #6)\par}}
\def\rfprocp#1;#2;#3;#4;#5;#6;#7 {{\frenchspacing\par\rn#1 #2, in {\it #3}, ed. #4 (#5: #6), p#7\par}}
\def\rfprep#1;#2;#3  {{\par\rn#1 #2, #3\par}}
\def\rfprepp#1;#2;#3 {{\par\rn#1 #2, #3\par}}

\def\kg{{\rm kg}}

\def\R{{\bf R}}

\def\etal{{\frenchspacing\it et al.}}
\def\ie{{\frenchspacing\it i.e.}}
\def\eg{{\frenchspacing\it e.g.}}
\def\etc{{\frenchspacing\it etc.}}

\def\beq#1{\begin{equation}\label{#1}}
\def\eeq{\end{equation}}
\def\beqa#1{\begin{eqnarray}\label{#1}}
\def\eeqa{\end{eqnarray}}

\def\sec#1{Section~\ref{#1}}

\def\spose#1{\hbox to 0pt{#1\hss}}
\def\simlt{\mathrel{\spose{\lower 3pt\hbox{$\mathchar"218$}}
     \raise 2.0pt\hbox{$\mathchar"13C$}}}
\def\simgt{\mathrel{\spose{\lower 3pt\hbox{$\mathchar"218$}}
     \raise 2.0pt\hbox{$\mathchar"13E$}}}
\def\simpropto{\mathrel{\spose{\lower 3pt\hbox{$\mathchar"218$}}
     \raise 2.0pt\hbox{$\propto$}}}

\def\ed{\end{document}}

\def\etal{{\frenchspacing\it et al.}}
\def\ie{{\frenchspacing\it i.e.}}
\def\eg{{\frenchspacing\it e.g.}}
\def\etc{{\frenchspacing\it etc.}}

\def\beq#1{$$}
\def\eeq{$$}

\def\fig#1{Figure~\ref{#1}}
\def\Fig#1{Figure~\ref{#1}}

\def\sec#1{Section~\ref{#1}}

\documentstyle[prl,aps,epsf,floats]{revtex}
\draft
\begin{document}
\twocolumn[\hsize\textwidth\columnwidth\hsize\csname@twocolumnfalse\endcsname

\title{Parallel Universes}

\author{Max Tegmark}

\address{Dept. of Physics, Univ. of Pennsylvania, 
Philadelphia, PA 19104; max@physics.upenn.edu}

\date{January 23 2003.
}

To appear in {\it Science and Ultimate Reality: From Quantum to Cosmos}, honoring John Wheeler's 90th birthday,
J.D.~Barrow, P.C.W.~Davies, \& C.L.~Harper eds., Cambridge University Press (2003)
\bigskip
\bigskip
\maketitle

\begin{abstract}
\noindent{\bf Abstract:}
I survey physics theories involving parallel universes, which 
form a natural four-level hierarchy of multiverses allowing progressively greater diversity.
Level I: A generic prediction of inflation is an infinite ergodic universe,
which contains Hubble volumes realizing all initial conditions --- including
an identical copy of you about $10^{10^{29}}$m away.
Level II: In chaotic inflation, other thermalized regions may have different physical 
constants, dimensionality and particle content.
Level III: In unitary quantum mechanics, other branches of the wavefunction 
add nothing qualitatively new, which is ironic given
that this level has historically been the most controversial.
Level IV: Other mathematical structures give different fundamental equations of physics.
The key question is not whether parallel universes exist (Level I is the uncontroversial 
cosmological concordance model), but how many levels there are.
I discuss how multiverse models can be falsified and argue that there is a severe 
``measure problem'' that must be solved to make testable predictions at levels II-IV.
\end{abstract}
\bigskip
] 

\setcounter{secnumdepth}{2}

Is there another copy of you 
reading this article, 
deciding to put it aside without finishing this sentence while you
are reading on? 
A person living on a planet called Earth, with
misty mountains, 
fertile fields
and sprawling cities, 
in a solar system with eight other planets.
The life of this person has been identical to yours in 
every respect -- until now, that is, when your decision to read on
signals that your two lives are diverging. 

You probably find this idea strange and
implausible, and I must confess that this 
is my gut reaction too.
Yet it looks like we will just have to live with it, since the
simplest and most popular cosmological model today predicts that this person 
actually exists in a Galaxy about $10^{10^{29}}$ meters from here.
This does not even assume speculative modern physics, merely 
that space is infinite and rather uniformly filled with matter
as indicated by recent astronomical observations.
Your {\it alter ego} is simply a prediction of the so-called 
concordance model of cosmology, which agrees with all current observational evidence and is 
used as the basis for most calculations and simulations presented at cosmology conferences.
In contrast, alternatives such as a fractal universe, a closed universe and a 
multiply connected universe have been seriously challenged by observations.

The farthest you can observe is the 
distance that light has been able to travel during the 14 billion years since the big-bang 
expansion began. The most distant visible objects are now about $4\times 10^{26}$ meters 
away\footnote{After emitting 
the light that is now reaching us, 
the most distant things we can see have receded because of the cosmic expansion,
and are now about about 40 billion light years away.
},
and a sphere of this radius defines our observable universe, also called our {\it Hubble volume}, our 
{\it horizon volume} or simply our universe. Likewise, the universe of your above-mentioned twin is a sphere of the same size 
centered over there, none of which we can see or have any causal contact with yet.
This is the simplest (but far from the only) example of parallel universes.

By this very definition of ``universe'', one might expect the
notion that our observable universe is merely a small part of a 
larger ``multiverse'' to be forever in the domain
of metaphysics. 
Yet the epistemological borderline between physics and 
metaphysics is defined by whether a theory is experimentally
testable, not by whether it is weird or involves unobservable entities.
Technology-powered experimental breakthroughs 
have therefore expanded the frontiers of physics to incorporate ever more abstract
(and at the time counterintuitive) concepts such as
a round rotating Earth, an electromagnetic field, time slowdown at high speeds,
quantum superpositions, curved space and black holes.
As reviewed in this article, it is becoming increasingly clear that 
multiverse models grounded in 
modern physics can in fact be empirically testable, predictive and falsifiable.
Indeed, as many as four distinct types of parallel universes (Figure 1) have been discussed in the 
recent scientific literature, so that the key question is not whether there is a multiverse
(since Level I is rather uncontroversial), but rather how many levels it has.

\begin{figure}[tbp]
\vskip-1.5cm
\centerline{{\vbox{\hglue-0.5cm\epsfxsize=19.4cm\epsfbox{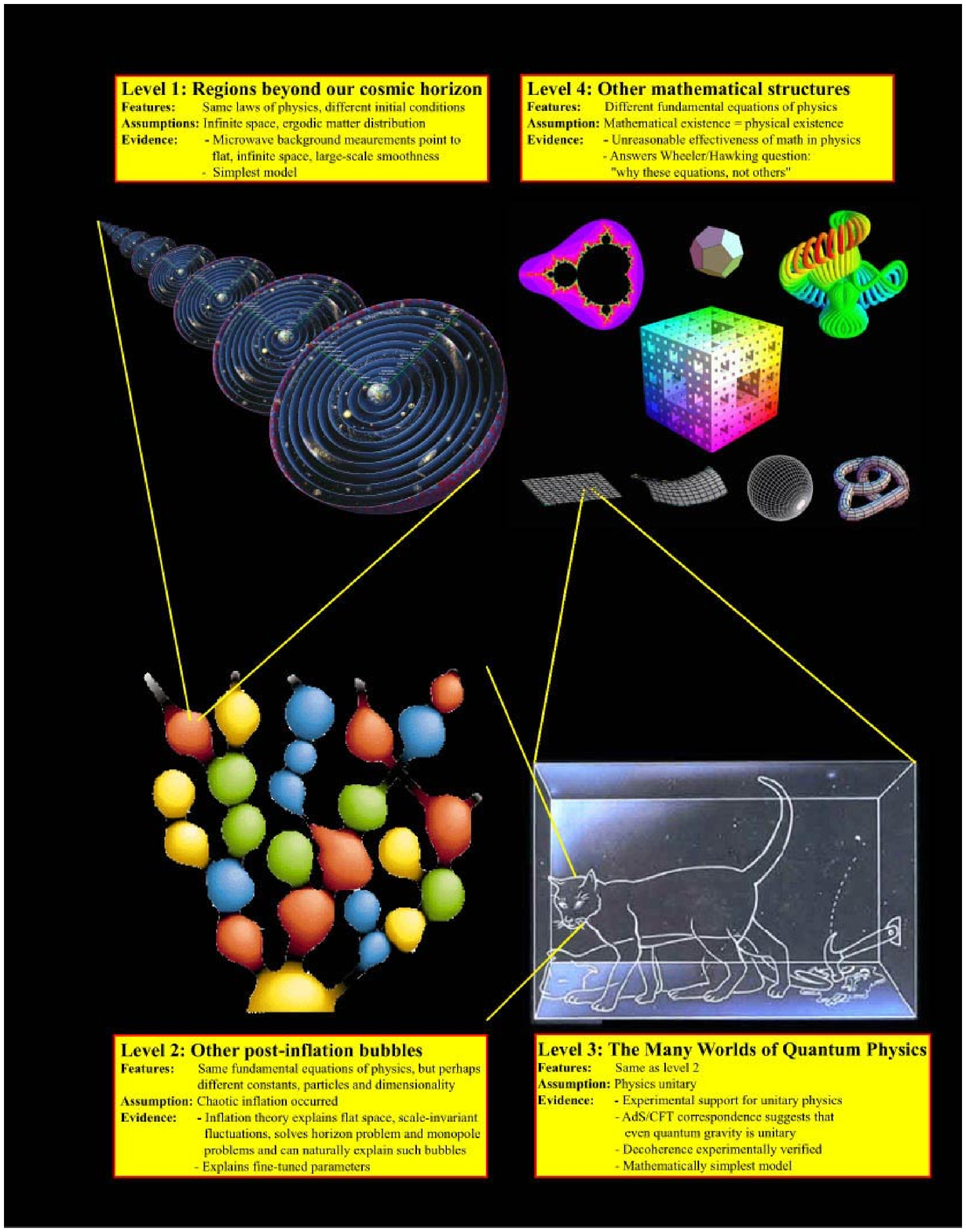}}}}
\label{ZoomFig}
\end{figure}
\setcounter{figure}{1}
\clearpage

\section{Level I: Regions beyond our cosmic horizon}

Let us return to your distant twin.
If space is infinite and the distribution of matter is sufficiently uniform on 
large scales, then even the most unlikely events must take place somewhere.
In particular, there are infinitely many other inhabited planets, including 
not just one but infinitely many with
people with the same appearance, name and memories as you.
Indeed, there are infinitely many other regions the size of our observable universe,
where every possible cosmic history is played out. This is the Level I multiverse.

\subsection{Evidence for Level I parallel universes}

Although the implications may seem crazy and counter-intuitive, this 
spatially infinite cosmological model is in fact the simplest and most popular one
on the market today. It is part of the cosmological concordance model, 
which agrees with all current observational evidence and is 
used as the basis for most calculations and simulations presented at cosmology conferences.
In contrast, alternatives such as a fractal universe, a closed universe and a multiply connected 
universe have been seriously challenged by observations.
Yet the Level I multiverse idea has been controversial (indeed, an assertion along these lines
was one of the heresies for which the Vatican had Giordano Bruno burned 
at the stake in 1600\footnote{Bruno's ideas have since been elaborated by, \eg,
Brundrit (1979), Garriga \& Vilenkin (2001b) and Ellis (2002), all of whom have
thus far avoided the stake.}), 
so let us review the status of the  two assumptions
(infinite space and ``sufficiently uniform'' distribution). 
 
\begin{figure}[pbt]
\centerline{{\vbox{\epsfxsize=8.7cm\epsfbox{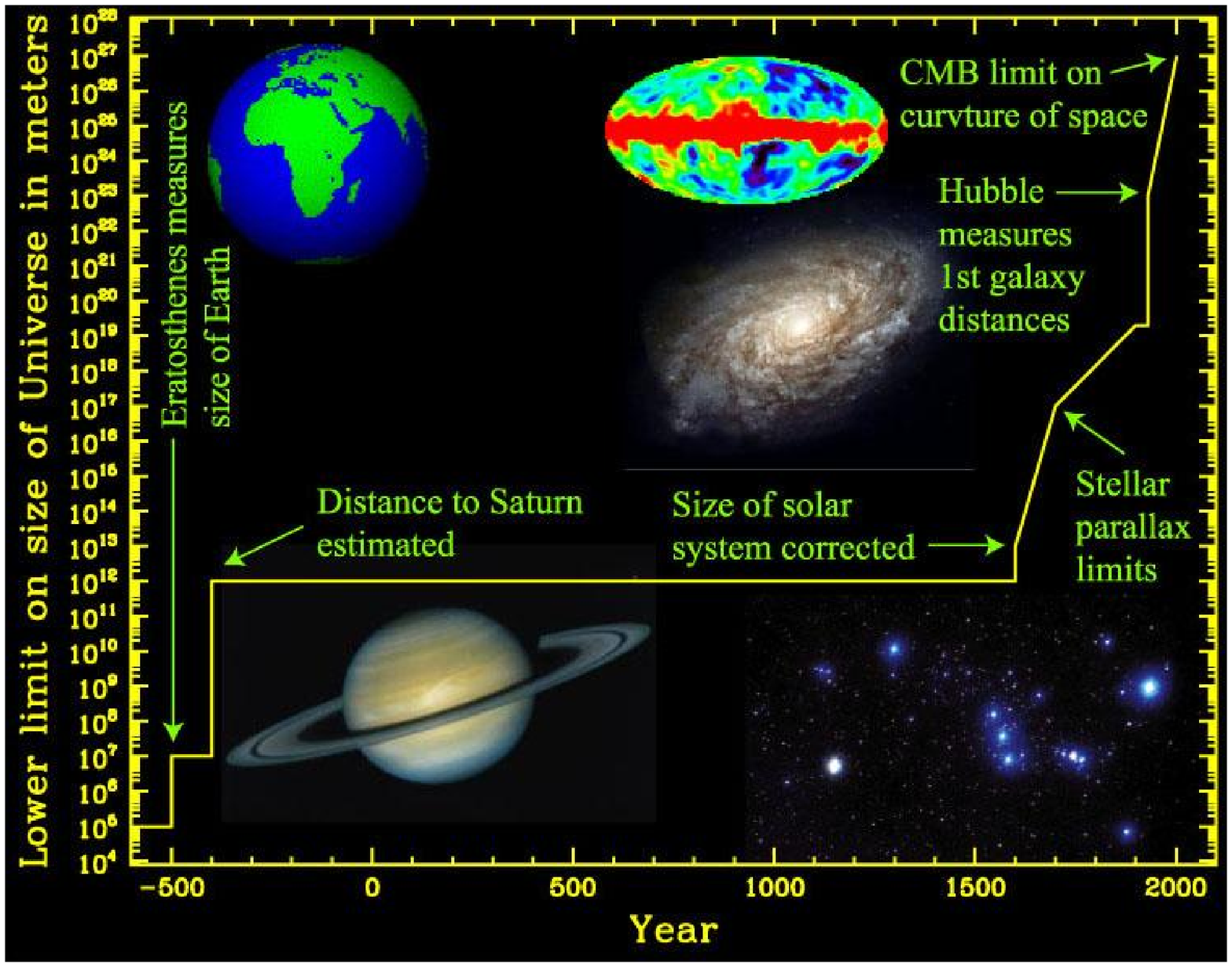}}}}
\smallskip
\caption{Although an infinite universe has always been a possibility, the lower
limit on the size of our universe has kept growing.
}
\label{SizeFig}
\end{figure}

How large is space? 
Observationally, the lower bound has grown dramatically 
(\fig{SizeFig}) with no indication of an upper bound.
We all accept the existence of things that we cannot see but could see
if we moved or waited, like ships beyond the horizon. 
Objects beyond cosmic horizon have similar status, since the observable universe 
grows by a light-year every year as light from further away has time to 
reach us\footnote{If the cosmic expansion continues to accelerate (currently an open question),
the observable universe will eventually stop growing.
}. 
Since we are all taught about simple Euclidean space in school, it can therefore
be difficult to imagine
how space could {\it not} be infinite --- for what would lie beyond the sign saying
{\it ``SPACE ENDS HERE --- MIND THE GAP''}?
Yet Einstein's theory of gravity allows space to be finite by
being differently connected than Euclidean space, say with the topology of 
a four-dimensional sphere
or a doughnut so that traveling far in one direction could bring
you back from the opposite direction.
The cosmic microwave background allows sensitive tests of such finite models,
but has so far produced no support for them --- flat infinite models 
fit the data fine and strong limits have been placed on both spatial curvature
and multiply connected topologies.
In addition, 
a spatially infinite universe is a generic prediction of the cosmological theory of 
inflation (Garriga \& Vilenkin 2001b).
The striking successes of inflation listed below therefore lend further support
to the idea that space is after all simple and infinite just as we learned in school.

How uniform is the matter distribution on large scales?
In an ``island universe'' model where space is infinite but all the matter is confined
to a finite region, almost all members of the Level I multiverse would be dead, consisting
of nothing but empty space. Such models have been popular historically, originally with the
island being Earth and the celestial objects visible to the naked eye, and in the early
20th century with the island being the known part of the Milky Way Galaxy.
Another non-uniform alternative is a fractal universe, where the matter distribution 
is self-similar and all coherent structures in the cosmic galaxy distribution are merely a 
small part of even larger coherent structures.
The island and fractal universe models have both been demolished by recent 
observations as reviewed in Tegmark (2002).
Maps of the three-dimensional galaxy distribution have shown that 
the spectacular large-scale structure observed (galaxy groups, clusters, superclusters, \etc)
gives way to dull uniformity on large scales,
with no coherent structures larger than about $10^{24}$m. 
More quantitatively, imagine placing a sphere of radius $R$ at various random
locations, measuring how much mass $M$ is enclosed each time, and computing the
variation between the measurements as quantified by their 
standard deviation $\Delta M$. The relative fluctuations
$\Delta M/M$ have been measured
to be of order unity on the scale 
$R\sim 3\times 10^{23}$m, and dropping on larger scales.
The Sloan Digital Sky Survey has found $\Delta M/M$ as small as  $1\%$ 
on the scale $R\sim 10^{25}$m and cosmic microwave background
measurements have established that the trend towards uniformity continues
all the way out to the edge of our observable universe ($R\sim 10^{27}$m), where
$\Delta M/M\sim 10^{-5}$.
Barring conspiracy theories where the universe is designed to fool us, the observations
thus speak loud and clear: space as we know it continues far beyond the edge of
our observable universe, teeming with galaxies, stars and planets.

\subsection{What are Level I parallel universes like?}
	\label{ErgodicitySec}

The physics description of the world is traditionally split into two parts:
initial conditions and laws of physics specifying how the initial conditions evolve.
Observers living in parallel universes at Level I observe the exact same laws of physics as we do,
but with different initial conditions than those in our Hubble volume.
The currently favored theory is that the initial conditions
(the densities and motions of different types of matter early on)
were created by quantum fluctuations during the inflation epoch (see section 3).
This quantum mechanism generates initial conditions that are for all practical 
purposes random, producing density fluctuations described by what mathematicians
call an ergodic random field.\footnote{Strictly speaking, the random field is ergodic if 
1) Space is infinite, 
2) the mass fluctuations $\Delta M/M$ approach zero on large scales (as measurements suggest),
and 
3) the densities at any set of points has a multivariate Gaussian probability distribution
(as predicted by the most popular inflation models, which can be traced back to the fact that
the harmonic oscillator equation governing the inflaton field fluctuations gives a 
Gaussian wavefunction for the ground state). 
For the technical reader, conditions 2 and 3 can be replaced by the weaker requirement that
correlation functions of all order vanish in the limit of infinite spatial separation.
}
{\it Ergodic} means that if you imagine 
generating an ensemble of universes, each with its own random initial conditions,
then the probability distribution of outcomes in a given volume
is identical to the distribution that you get by sampling different volumes in a single universe.
In other words, it means that everything that could in principle have happened here 
did in fact happen somewhere else.

Inflation in fact generates all possible initial conditions with non-zero probability, 
the most likely ones
being almost uniform with fluctuations at the $10^{-5}$ level that are amplified by
gravitational clustering to form galaxies, stars, planets and other structures.
This means both that pretty much all imaginable matter configurations occur in some Hubble 
volume far away, and also that we should expect our own Hubble volume to be a
fairly typical one --- at least typical among those that contain observers.
A crude estimate suggests that the closest identical copy of you is about $\sim10^{10^{29}}$m away. 
About $\sim 10^{10^{91}}$m away, there should be a sphere of radius 
100 light-years identical to the one centered here, so all perceptions that we have
during the next century will be identical to those of our counterparts over there.
About $\sim 10^{10^{115}}$m away, there should be an entire Hubble volume identical to 
ours.\footnote{This 
is an extremely conservative estimate, simply counting all possible quantum 
states that a Hubble volume can have that are no hotter than $10^8$K.  
$10^{115}$ is roughly the number of protons that the Pauli exclusion principle
would allow you to pack into a Hubble volume at this temperature
(our own Hubble volume contains only about $10^{80}$ protons).
Each of these $10^{115}$ slots can be either occupied or unoccupied, giving 
$N=2^{10^{115}}\sim 10^{10^{115}}$ possibilities, 
so the expected distance to the nearest identical Hubble volume is 
$N^{1/3}\sim 10^{10^{115}}$ Hubble radii $\sim 10^{10^{115}}$ meters.
Your nearest copy is likely to 
be much closer than $10^{10^{29}}$ meters, since the planet formation and evolutionary processes 
that have tipped the odds in your favor are at work everywhere. There are probably at 
least $10^{20}$ habitable planets in our own Hubble volume alone. 
}

This raises an interesting philosophical point that will come back and haunt us in \sec{MeasureSec}: 
if there are indeed many copies of ``you'' with identical past lives and memories, 
you would not be able to compute your own future even if you had complete knowledge
of the entire state of the cosmos! 
The reason is that there is no way for you to determine 
which of these copies is ``you'' (they all feel that they are). Yet their lives will typically 
begin to differ eventually, so the best you can do is predict probabilities for what you will
experience from now on.  
This kills the traditional notion of determinism.

\subsection{How a multiverse theory can be tested and falsified}
\label{TestingSec}

Is a multiverse theory one of metaphysics rather than physics?
As emphasized by Karl Popper, the distinction between the two is whether the theory
is empirically testable and falsifiable.
Containing unobservable entities does clearly {\it not} per se make a theory non-testable.
For instance, a theory stating that there are 666 parallel universes, 
all of which are devoid of oxygen makes the testable prediction that we should observe no oxygen here, 
and is therefore ruled out by observation.

As a more serious example, the Level I multiverse framework is routinely used 
to rule out theories in modern cosmology, although this is rarely spelled out explicitly. 
For instance, cosmic microwave background (CMB) observations have recently 
shown that space has almost no curvature.
Hot and cold spots in CMB maps have a characteristic 
size that depends on the curvature of space, and the observed spots appear
too large to be consistent with the previously popular ``open universe'' model.
However, the average spot size randomly varies slightly from one Hubble 
volume to another, so it is important to be statistically rigorous.
When cosmologists say that the open universe model is ruled out at 
99.9\% confidence,
they really mean that if the open universe model were true,
then fewer than one out of every thousand Hubble volumes would show CMB spots as
large as those we observe --- therefore the entire model with all its infinitely many 
Hubble volumes is ruled out, even though we have of course only 
mapped the CMB in our own particular Hubble volume.

The lesson to learn from this example is that multiverse theories {\it can} be 
tested and falsified, but only if they predict 
what the ensemble of parallel universes is and specify  a probability distribution
(or more generally what mathematicians call a {\it measure})
over it. As we will see in \sec{MeasureSec}, 
this measure problem can be quite serious and is still unsolved for
some multiverse theories.

\section{Level II: Other post-inflation bubbles}

If you felt that the Level I multiverse was large and hard to stomach,
try imagining an infinite set of distinct ones
(each symbolized by a bubble in 
Figure 1), some perhaps with different
dimensionality and different physical constants.
This is what is predicted by
the the currently popular chaotic theory of inflation, and we will refer to
it as the Level II multiverse. 
These other domains are more than infinitely far away in the sense
that you would never get there even if you traveled at the speed of light forever.
The reason is that the space between our Level I multiverse and its neighbors
is still undergoing inflation, which keeps stretching it out 
and creating more volume faster than you can travel through it.
In contrast, you could travel to an arbitrarily distant Level I universe if you were patient 
and the cosmic expansion decelerates.\footnote{
Astronomical evidence suggests that the cosmic expansion is currently accelerating.
If this acceleration continues, then even the level I parallel universes will remain forever separate, 
with the intervening space stretching faster than light can travel through it.
The jury is still out, however, with popular models predicting that the universe will eventually 
stop accelerating and perhaps even recollapse.
}

\subsection{Evidence for Level II parallel universes}

By the 1970's, the Big Bang model had proved a highly successful explanation 
of most of the history of our universe. It had explained how a primordial fireball 
expanded and cooled, synthesized Helium and other
light elements during the first few minutes, became transparent after 400,000 years releasing
the cosmic microwave background radiation, and gradually got clumpier due to gravitational 
clustering, producing galaxies, stars and planets.
Yet disturbing questions remained about what happened in the very beginning.
Did something appear from nothing?
Where are all the superheavy particles known as magnetic monopoles that particle physics
predicts should be created early on (the ``monopole problem'')?
Why is space so big, so old and so flat, when generic initial conditions predict
curvature to grow over time and the density to approach either zero or infinity after
of order $10^{-42}$ seconds (the ``flatness problem'')?
What conspiracy caused the CMB temperature to be nearly identical in regions
of space that have never been in causal contact (the ``horizon problem'')?
What mechanism generated the $10^{-5}$ level seed fluctuations 
out of which all structure grew?

A process known as  {\it inflation} can solve all these problems 
in one fell swoop (see reviews by Guth \& Steinhardt 1984 and Linde 1994),
and has therefore emerged as the 
most popular theory of what happened very early on.
Inflation is a rapid stretching of space, diluting away monopoles and other debris, 
making space flat and uniform like the surface of an expanding balloon, 
and stretching quantum vacuum fluctuations into macroscopically large density fluctuations
that can seed galaxy formation. Since its inception, inflation has passed additional tests:
CMB observations have found space to be extremely flat
and have measured the seed fluctuations to have an approximately scale-invariant spectrum
without a substantial gravity wave component, all in perfect agreement with 
inflationary predictions.

Inflation is a general phenomenon that occurs in 
a wide class of theories of elementary particles.
In the popular model known as {\it chaotic inflation},
inflation ends in some regions of space allowing life as we know it, whereas 
quantum fluctuations cause other regions of space to inflate even faster.
In essence, one inflating bubble sprouts other inflationary bubbles, which in 
turn produce others in a never-ending chain reaction 
(Figure 1, lower left, with time increasing upwards).
The bubbles where inflation has ended are the elements of the Level II multiverse.
Each such bubble is infinite 
in size\footnote{Surprisingly, it has been shown that inflation can produce 
an infinite Level I multiverse even in a bubble of finite spatial volume, thanks
to an effect whereby the spatial directions of spacetime curve towards the (infinite) 
time direction (Bucher \& Spergel 1999). 
}, 
yet there are infinitely many bubbles since the chain reaction never ends. 
Indeed, if this exponential growth of the number of bubbles 
has been going on forever, there will be an uncountable 
infinity of such parallel universes
(the same infinity as that assigned to the set of real numbers, say,
which is larger than that of the [countably infinite] set
of integers).
In this case, there is also no beginning of time and no absolute Big Bang: 
there is, was and always will be an infinite number of inflating bubbles and post-inflationary
regions like the one we inhabit, forming a fractal pattern.

\subsection{What are  Level II parallel universes like?}

The prevailing view is that the physics we observe today is
merely a low-energy limit of a much more symmetric theory that 
manifests itself at extremely high temperatures. This underlying fundamental theory
may be 11-dimensional, supersymmetric and involving a grand unification 
of the four fundamental forces of nature.
A common feature in such theories is that the potential energy of the
field(s) driving inflation has several different minima 
(sometimes called ``vacuum states''), 
corresponding to different ways of breaking this symmetry and, 
as a result, to different low-energy physics.
For instance, all but three spatial dimensions could be curled up (``compactified''), 
resulting in an effectively three-dimensional space like ours, or 
fewer could curl up leaving a 7-dimensional space.
The quantum fluctuations driving chaotic inflation could cause different symmetry breaking 
in different bubbles, resulting in different members of the Level II multiverse having different 
dimensionality.
Many symmetries observed in particle physics also result from the specific way in
which symmetry is broken, so there could be Level II parallel universes where there
are, say, two rather than three generations of quarks. 

In addition to such discrete properties as dimensionality and fundamental particles,
our universe is characterized by a set of dimensionless 
numbers known as {\it physical constants}. Examples include the electron/proton mass ratio $m_p/m_e\approx 1836$ and the cosmological constant, which appears to be about
$10^{-123}$ in so-called Planck units. There are models where also such 
continuous parameters can vary from one one post-inflationary 
bubble to another.\footnote{
Although the fundamental equations of physics are the same throughout the Level II multiverse, the 
approximate effective equations governing the low-energy world that we observe will differ.
For instance, moving from a three-dimensional to a four-dimensional (non-compactified) space 
changes the observed gravitational force equation from an inverse square law to an inverse cube law.
Likewise, breaking the underlying symmetries of particle physics differently will change the lineup 
of elementary particles and the effective equations that describe them.
However, we will reserve the terms ``different equations'' and ``different laws of physics''
for the Level IV multiverse, where it is the fundamental rather than effective equations that change.
}

The Level II multiverse is therefore likely to be more diverse than the Level I multiverse,
containing domains where not only the initial conditions differ,
but perhaps the dimensionality, the elementary particles and
the physical constants differ as well.

Before moving on, let us briefly comment on a few closely related multiverse notions.
First of all, if one Level II multiverse can exist, eternally self-reproducing in a fractal pattern,
then there may well be infinitely many other Level II multiverses that are completely disconnected.
However, this variant appears to be untestable, since it would neither add any qualitatively different worlds 
nor alter the probability distribution for their properties.
All possible initial initial conditions and symmetry breakings
are already realized within each one.

An idea proposed by Tolman and Wheeler and recently elaborated by
Steinhardt \& Turok (2002) is that the (Level I) multiverse is
cyclic, going through an infinite series of Big Bangs.
If it exists, the ensemble of such incarnations would also form
a multiverse, arguably with a diversity similar to that of Level II.

An idea proposed by Smolin (1997) involves an ensemble similar in diversity to that of Level II, but 
mutating and sprouting new universes through black holes rather than during inflation.
This predicts a form of a natural selection favoring universes with maximal black hole production.

In braneworld scenarios, another 3-dimensional world could be quite
literally parallel to ours, merely offset in a higher dimension. However, it is unclear
whether such a world (``brane'') deserves be be called a parallel universe separate from our own, since 
we may be able to interact with it gravitationally much as we do with dark matter.

\begin{figure}[pbt]
\vskip-0.4cm
\centerline{{\vbox{\epsfxsize=8.8cm\epsfbox{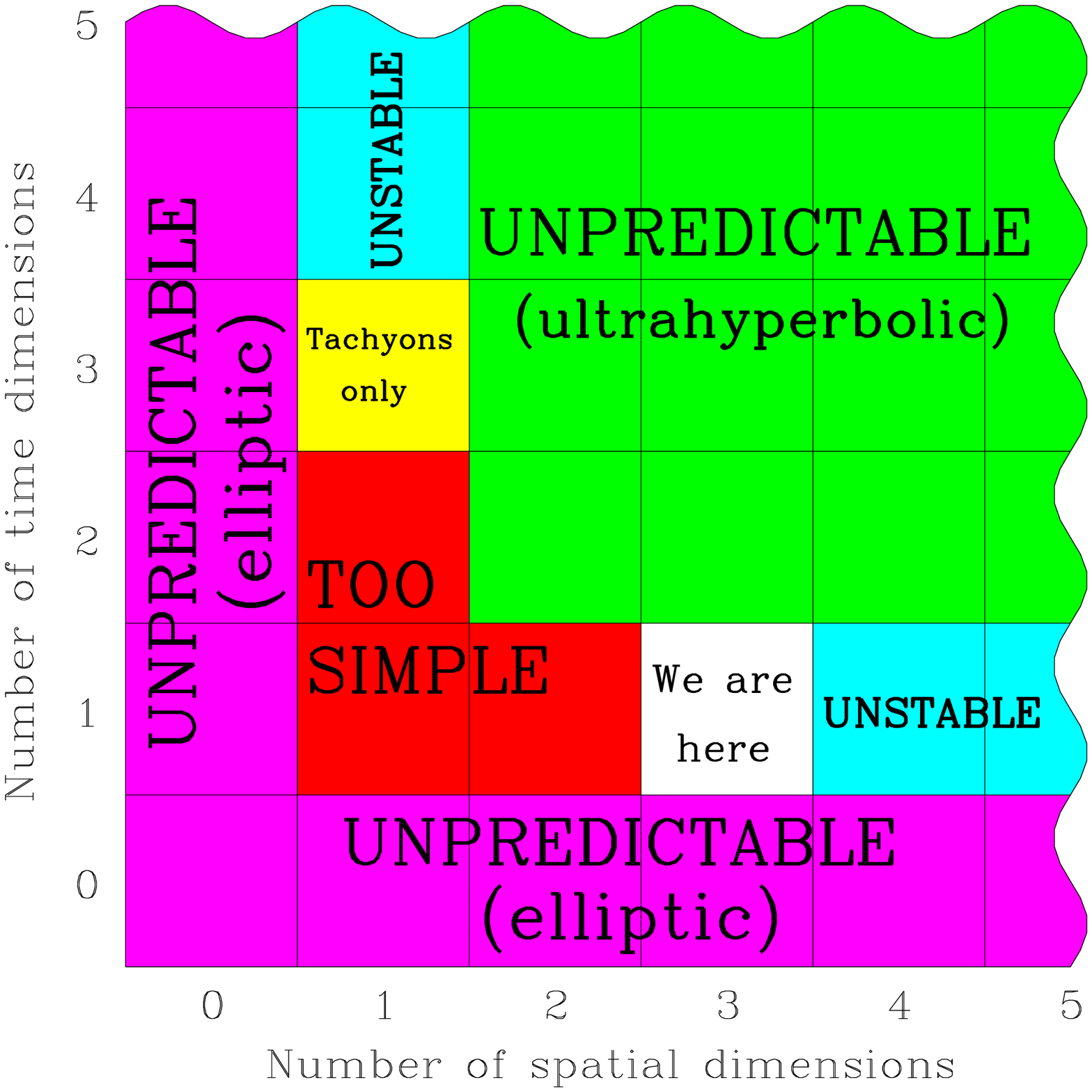}}}}
\smallskip
\caption{
\label{DimensionsFig} 
Why we should not be surprised to find ourselves living in 3+1-dimensional spacetime.
When the partial differential equations of nature are elliptic or
ultrahyperbolic, physics has no predictive power for an observer.
In the remaining (hyperbolic)
cases, $n>3$ admits no stable atoms and $n<3$ may lack sufficient complexity for observers
(no gravitational attraction, topological problems). From Tegmark (1997).
}
\vskip-0.1cm
\end{figure}

\subsection{Fine-tuning and selection effects}
\label{TuningSec}

Physicists dislike unexplained coincidences.
Indeed, they interpret them as evidence that models are ruled out.
In \sec{TestingSec}, we saw how the open universe model was 
ruled out at 99.9\% confidence because it implies
that the observed pattern of CMB fluctuations is extremely unlikely, 
a one-in-a thousand coincidence occurring in only 0.1\% of all Hubble volumes.

Suppose you check into a hotel, are assigned room 1967 and, surprised, note that 
that this is the year you were born. After a moment of reflection, you conclude that
this is not all that surprising after all, given that  the hotel has many rooms and that
you would not be having these thoughts in the first place if you'd been assigned 
another one. You then realize that even if you knew nothing about 
hotels, you could have inferred the existence of other hotel rooms, because
if there were  only one room number in the entire universe, you would be
left with an unexplained coincidence.

As a more pertinent example, consider 
$M$, the mass of the Sun.
$M$ affects the luminosity of the Sun, and 
using basic physics, one can compute that life as we know it on Earth 
is only possible if $M$ is in the narrow range $1.6\times 10^{30}\kg - 2.4\times 10^{30}\kg$ --- otherwise 
Earth's climate would be colder than on Mars or hotter than on Venus.
The measured value  is $M\sim 2.0\times 10^{30}\kg$.
This apparent coincidence of 
the habitable and observed $M$-values may appear disturbing given that 
calculations show that stars in the much broader 
mass range $M\sim 10^{29}\kg-10^{32}\kg$ can exist. 
However, just as in the hotel example, we can explain this apparent coincidence 
if there is an ensemble and a selection effect: 
if there are in fact many solar systems with a range of sizes of the central star and the 
planetary orbits, then we obviously expect to find ourselves living in one of
the inhabitable ones. 

More generally, the apparent coincidence of the 
habitable and observed values of some physical parameter can be taken as evidence 
for the existence of a larger ensemble, of which what we observe is merely one
member among many (Carter 1973).
Although the existence of other hotel rooms and solar systems is uncontroversial 
and observationally confirmed,
that of parallel universes is not, since they cannot be observed.
Yet if fine-tuning is observed, one can argue for their existence using the exact same 
logic as above.
Indeed, there are numerous examples of fine tuning suggesting 
parallel universes with other physical constants, although the degree of fine 
tuning is still under active debate and should be clarified by additional calculations
--- see Rees (2002) and Davies (1982) for popular accounts and 
Barrow \& Tipler (1986) for technical details.

For instance, if the electromagnetic force were weakened by a mere $4\%$,
then the Sun would immediately explode
(the diproton would have a bound state, which would increase the solar
luminosity by a factor $10^{18}$). 
If it were stronger, there would be fewer stable atoms.
Indeed, most if not all the parameters affecting low-energy physics appear fine-tuned
at some level, in the sense that changing them by modest amounts results in 
a qualitatively different universe.

\begin{figure}[pbt]
\vskip-0.3cm
\centerline{{\vbox{\epsfxsize=8.8cm\epsfbox{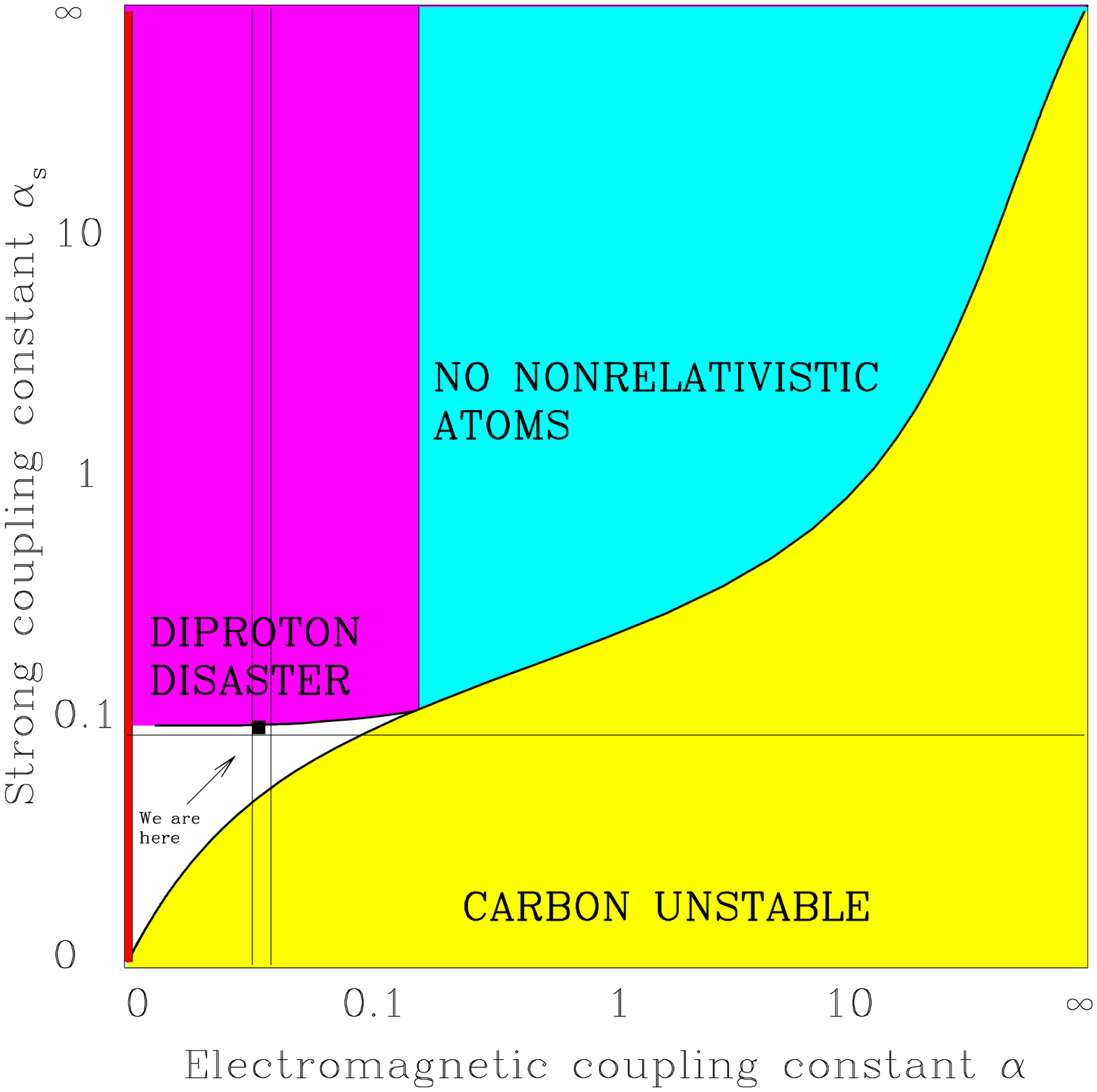}}}}
\smallskip
\caption{
Hints of fine-tuning for the parameters $\alpha$ and $\alpha_s$ which
determine the strengths of the 
electromagnetic force and the strong nuclear force, respectively
(from Tegmark 1997).
The observed values $(\alpha,\alpha_s)\approx(1/137,0.1)$
are indicated with a filled square. Grand unified theories
rule out everything except the narrow strip between the two vertical lines,
and deuterium becomes unstable below the horizontal line.
In the narrow shaded region to the very left,
electromagnetism is weaker than gravity and
therefore irrelevant.
\label{alphaalphasFig} 
}
\vskip-0.6cm
\end{figure}

If the weak interaction were substantially weaker, there would be no hydrogen around, 
since it would have been 
converted to helium shortly after the Big Bang.
If it were either much stronger or much weaker, the neutrinos from a supernova
explosion would fail to blow away the outer parts of the star,
and it is doubtful whether life-supporting heavy elements would
ever be able to leave the stars where they were produced.
If the protons were $0.2\%$ 
heavier, they would decay into neutrons unable to hold onto electrons, so there would be
no stable atoms around.
If the proton-to-electron mass ratio were much smaller, there could be no stable stars,
and if it were much larger, there could be no ordered structures like crystals
and DNA molecules.

Fine-tuning discussions often turn heated when somebody mentions the
``A-word'', {\it anthropic}. The author feels that discussions of the so-called anthropic 
principle have generated more heat than light, with many different definitions and interpretations
of what it means.
The author is not aware of anybody disagreeing with 
what might be termed MAP, the {\it minimalistic anthropic principle}:
\begin{itemize}
\item {\it MAP: When testing fundamental theories with observational data, 
ignoring selection effects can give incorrect conclusions.}
\end{itemize}
This is obvious from our examples above: if we neglected selection 
effects, we would be surprised to orbit a star as heavy as the Sun, since 
lighter and dimmer ones are much more abundant. 
Likewise, MAP says that the chaotic inflation model is {\it not} ruled out
by the fact that we find ourselves living in the minuscule fraction of space where inflation has ended,
since the inflating part is uninhabitable to us.
Fortunately, selection effects cannot rescue all models, as pointed out a
century ago by 
Boltzmann.
If the universe were in classical thermal equilibrium (heat death), thermal fluctuations
could still make atoms assemble at random to briefly create a self-aware observer like you
once in a blue moon, so the fact that you exist right now does not rule out the heat death cosmological 
model. However, you should statistically expect to find the
rest of the world in a high-entropy mess rather than in the ordered low-entropy state you observe, 
which rules out this model.

The standard model of particle physics has 28 of free parameters, and cosmology may 
introduce additional independent ones. 
If we really do live in a Level II multiverse, then for those parameters that vary
between the parallel universes,
we will never be able to predict our measured values from first principles. We can merely compute 
probability distributions for what we should expect to find, taking selection effects into account.
We should expect to find everything that can vary across the ensemble to be
as generic as is consistent with our existence.
As detailed in \sec{MeasureSec}, 
this issue of what is ``generic'' and, more specifically, how to compute probabilities in physics,
is emerging as an embarrassingly thorny problem (see \sec{MeasureSec}).

\begin{figure}[pbt]
\centerline{{\vbox{\epsfxsize=8.7cm\epsfbox{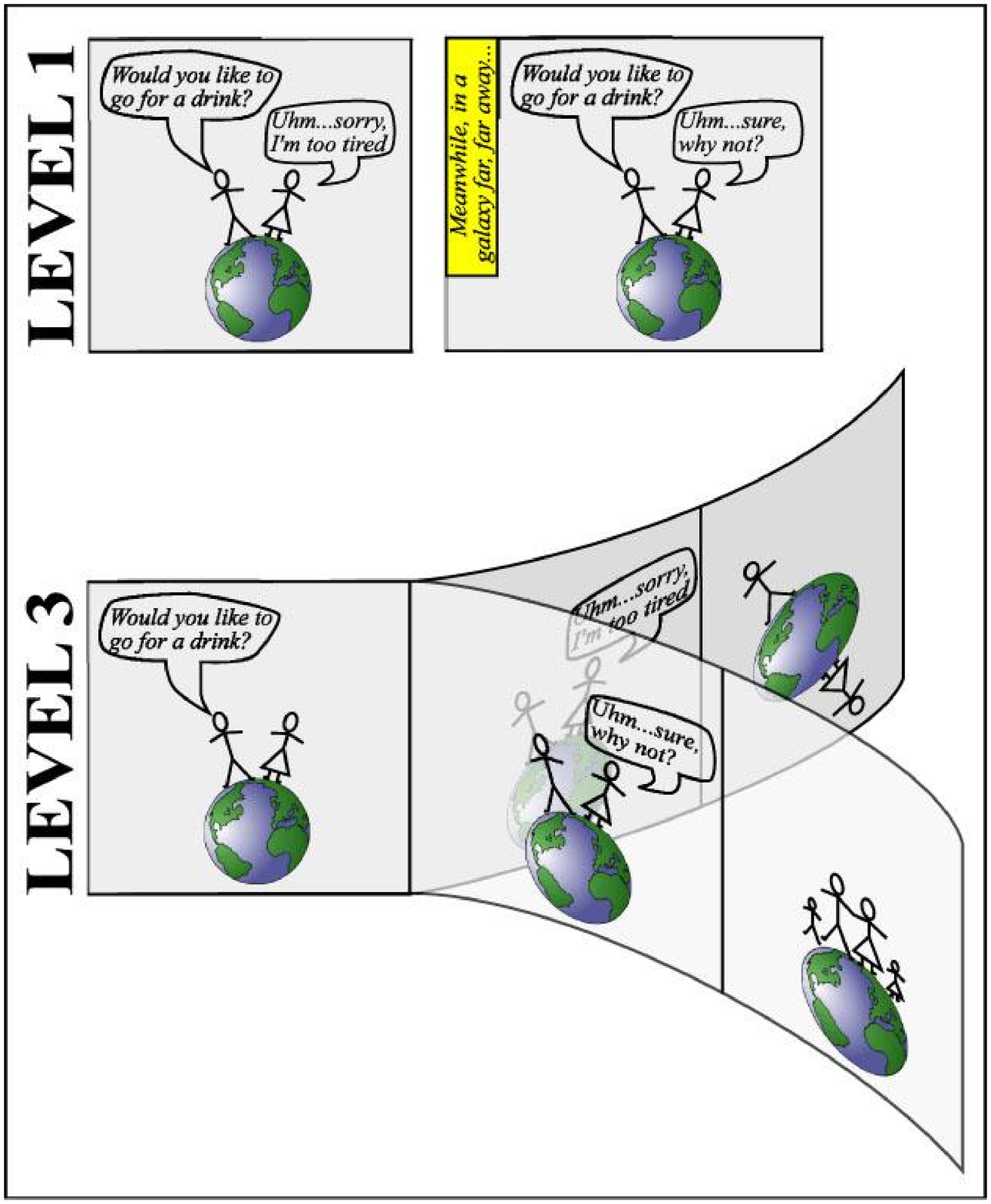}}}}
\smallskip
\caption{
Difference between Level I and Level III.
Whereas Level I parallel universes are far away in space, those of Level III
are even right here, with quantum events causing classical reality to split and diverge into 
parallel storylines. Yet Level III adds no new storylines beyond levels 1 or 2.
}
\label{CartoonFig}
\end{figure}

\section{Level III: The many worlds of quantum physics}

There may be a third type of parallel worlds that are 
not far away but in a sense right here.
If the fundamental equations of physics are what mathematicians call {\it unitary}, 
as they so far appear to be, then the universe keeps branching into parallel universes 
as in the cartoon (\fig{CartoonFig}, bottom): whenever a quantum event appears to have a random outcome, 
all outcomes in fact occur, one in each branch.
This is the Level III multiverse. Although more debated and controversial than 
Level I and Level II, we will see that, surprisingly, 
this level adds no new types of universes.

\subsection{Evidence for Level III parallel universes}

In the early 20th century, the theory of quantum mechanics revolutionized physics
by explaining the atomic realm, with applications ranging from chemistry to
nuclear reactions, lasers and semiconductors.
Despite the obvious successes in its application, a heated debate ensued about its
interpretation --- a debate that still rages on.
In quantum theory, the state of the universe is not given in classical terms
such as the positions and velocities of all particles, but by a mathematical
object called a wavefunction. According to the so-called Schr\"odinger equation,
this state evolves deterministically over time in a fashion termed 
{\it unitary}, corresponding to a rotation in Hilbert space, the abstract infinite-dimensional 
space where the wavefunction lives.
The sticky part is that there are perfectly legitimate wavefunctions corresponding
to classically counterintuitive situations such as you being in two different 
places at once. Worse, the Schr\"odinger equation can evolve innocent classical states
into such schizophrenic ones.
As a baroque example, Schr\"odinger described the famous thought experiment where a
nasty contraption kills a cat if a radioactive atom decays. Since the
radioactive atom eventually enters a superposition of decayed and not
decayed, it produces a cat which is both dead and alive in
superposition.

In the 1920s, this weirdness was explained away by postulating that
that the wavefunction ``collapsed'' into some definite classical outcome 
whenever an observation was made, with probabilities given by the wavefunction.
Einstein was unhappy about such intrinsic randomness in nature, which violated unitarity,
insisting that ``God doesn't play dice'', and others complained that there was no equation
specifying when this collapse occurred.
In his 1957 Ph.D. thesis, Princeton student Hugh Everett III 
showed that this controversial collapse postulate was unnecessary.
Quantum theory predicted
that one classical reality would gradually split into superpositions of 
many (\fig{CartoonFig}). He showed that observers would subjectively
experience this splitting merely as a slight randomness, and indeed with probabilities 
in exact agreement with those from the old collapse postulate (de Witt 2003).
This superposition of classical worlds is the Level III multiverse.

Everett's work had left two crucial questions unanswered: first of all, if the world
actually contains bizarre macrosuperpositions, then why don't we perceive
them?
The answer came in 1970, when Dieter Zeh
showed that the Schr\"odinger equation itself gives rise to a type of censorship
effect (Zeh 1970). This effect became known as {\it decoherence}, and was worked out in
great detail by Wojciech Zurek, Zeh and others over the following
decades. Coherent quantum superpositions were found to persist only as
long as they were kept secret from the rest of the world.
A single collision with a snooping photon or air molecule is sufficient to ensure that our
friends in \fig{CartoonFig}  can never be aware  of their counterparts in the parallel storyline.
A second unanswered question in the Everett picture was more subtle but
equally important: what physical mechanism picks out approximately
classical states (with each object in only one place, \etc) 
as special in the bewilderingly large Hilbert space?
Decoherence answered this question as well,
showing that classical states are simply those that are most robust 
against decoherence. In summary, decoherence both identifies the Level III parallel 
universes in Hilbert space and delimits them from one another.
Decoherence is now quite uncontroversial and 
has been experimentally measured in a wide range of circumstances.
Since decoherence for all practical purposes mimics wavefunction collapse, 
it has eliminated much of the original motivation for non-unitary quantum mechanics
and made the Everett's so-called many worlds interpretation increasingly popular.
For details about these quantum issues, see Tegmark \& Wheeler (2001) for a popular account
and Giulini {\etal} (1996) for a technical review.

If the time-evolution of the wavefunction is unitary, then the Level III multiverse exists,
so physicists have worked hard on testing this crucial assumption.
So far, no departures from unitarity have been found.
In the last few decades, remarkable experiments have confirmed
unitarity for ever larger systems, including the hefty carbon-60 ``Buckey Ball'' atom
and kilometer-size optical fiber systems.
On the theoretical side, a leading argument against unitarity has involved possible 
destruction of information during the evaporation of black holes, suggesting that
quantum-gravitational effects are non-unitary and collapse the wavefunction.
However, a recent string theory breakthrough known as
AdS/CFT correspondence has suggested that even quantum gravity is unitary,
being mathematically equivalent to a lower-dimensional quantum field theory without gravity
(Maldacena 2003).

\subsection{What are Level III parallel universes like?}

When discussing parallel universes, we need to distinguish between two 
different ways of viewing a physical theory: 
the outside view or {\it bird perspective} of a mathematician 
studying its mathematical fundamental equations and the 
inside view or {\it frog perspective} of an observer living in the
world described by the 
equations\footnote{
Indeed, the standard mental picture of what the physical world is corresponds  
to a third intermediate viewpoint that could be termed the 
{\it consensus view}.
From your subjectively perceived frog perspective, the world turns upside down when you stand on your head
and disappears when you close your eyes, yet you subconsciously interpret your sensory inputs 
as though there is an external reality that is 
independent of your orientation, your location and your state of mind.
It is striking that although this third view involves both
censorship (like rejecting dreams),
interpolation (as between eye-blinks)
and extrapolation (say attributing existence to unseen cities)
of your inside view, independent observers nonetheless appear to share this 
consensus view.
Although the inside view looks black-and-white to a cat, 
iridescent to a bird seeing four primary colors,
and still more different to bee a seeing polarized light, a bat using sonar,
a blind person with keener touch and hearing, or the latest overpriced
robotic vacuum cleaner, all agree on whether the door is open.
The key current challenge in physics is deriving this semiclassical 
consensus view from the fundamental equations specifying the bird perspective.
In my opinion, this means that 
although understanding the detailed nature of human consciousness is an
important challenge in its own right, it is {\it not} necessary 
for a fundamental theory of physics.
}.
From the bird perspective, the Level III multiverse is simple: 
there is only one wavefunction, and it evolves smoothly and deterministically over time
without any sort of splitting or parallelism. 
The abstract quantum world described by this evolving wavefunction contains within it
a vast number of parallel classical storylines (see \fig{CartoonFig}), continuously splitting and merging,  as well as a number of quantum phenomena that lack a classical description.
From her frog perspective, however, each observer perceives only a tiny fraction of this full reality:
she can only see her own Hubble volume (Level I) and decoherence prevents her from perceiving
Level III parallel copies of herself. When she is asked a question, makes 
a snap decision and answers (\fig{CartoonFig}), quantum effects at the neuron level in her brain 
lead to multiple outcomes, and from the bird perspective, her single past branches into multiple futures.
From their frog perspectives, however, each copy of her is unaware of the other copies, and 
she perceives this quantum branching as merely a slight randomness.
Afterwards, there are for all practical purposes multiple copies of her that have the exact same 
memories up until the point when she answers the question.

\subsection{How many different parallel universes are there?}
As strange as this may sound, \fig{CartoonFig} illustrates that this exact same situation occurs even in
the Level I multiverse, the only difference being where her copies reside 
(elsewhere in good old three-dimensional space as opposed to 
elsewhere in infinite-dimensional Hilbert space, in other quantum branches).
In this sense, Level III is no stranger than Level I. 
Indeed, if physics is unitary, then the quantum fluctuations during inflation
did not generate unique initial conditions through a random process, 
but rather generated a quantum superposition
of all possible initial conditions simultaneously, after which decoherence 
caused these fluctuations to behave essentially classically in separate quantum branches. 
The ergodic nature of these quantum fluctuations (\sec{ErgodicitySec})
therefore implies that the distribution of outcomes 
in a given Hubble volume at Level III (between different quantum branches as in Fig 3)
is identical to the distribution that you get by sampling different Hubble volumes within a single 
quantum branch (Level I). 
If physical constants, spacetime dimensionality {\etc} can vary as in Level II,
then they too will vary between parallel quantum branches at Level III. The reason for
this is that if physics is unitary, then the process of spontaneous symmetry breaking 
will not produce a unique (albeit random) outcome, but rather a superposition of
all outcomes that rapidly decoheres into for all practical purposes separate 
Level III branches.
In short, the Level III multiverse, if it exists, adds nothing new 
beyond Level I and Level II --- just more indistinguishable copies of the same universes, the
same old storylines playing out again and again in other quantum branches.
Postulating a yet unseen non-unitary effect to get rid of the Level III multiverse, 
with Ockham's Razor in mind, therefore would not make Ockham any happier.

The passionate debate about Everett's parallel universes that has raged on for
decades therefore seems to be ending in a grand anticlimax, with the 
discovery of a less controversial multiverse that is just as large.
This is reminiscent of the famous Shapley-Curtis debate of the 1920s about
whether there were really a multitude of galaxies 
(parallel universes by the standards of the time) or just one, a storm in a teacup 
now that research has moved on to other galaxy clusters, superclusters and even Hubble volumes.
In hindsight, both the Shapley-Curtis and Everett controversies seem positively quaint,
reflecting our instinctive reluctance to expand our horizons.

\begin{figure}[pbt]
\centerline{{\vbox{\epsfxsize=8.5cm\epsfbox{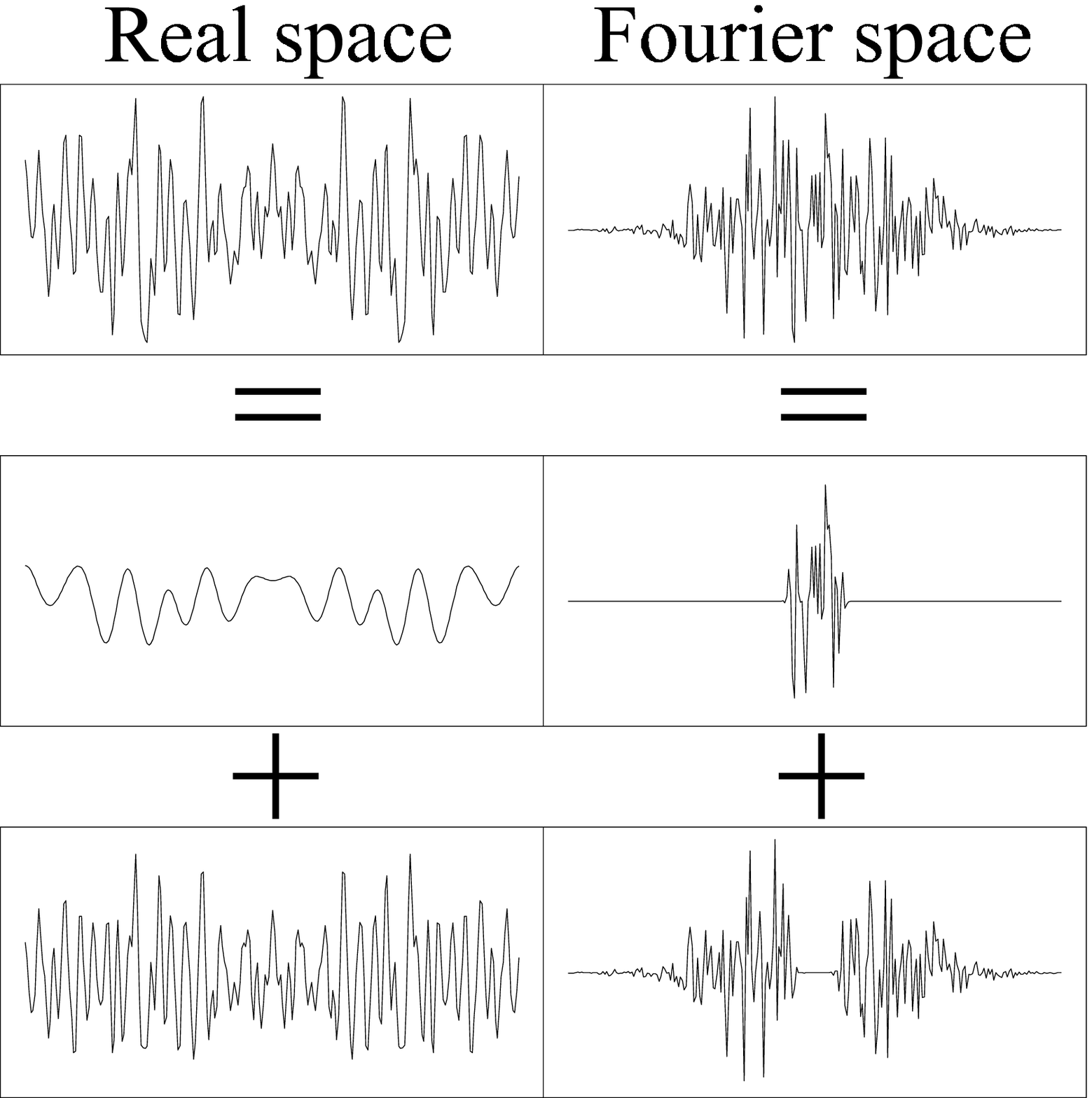}}}}
\smallskip
\caption{
Schematic illustration (see footnote) of how a wavefunctional of the Level 3 multiverse
(top row for simple 1-dimensional Hilbert space)
can be decomposed as a superposition of semiclassical worlds 
(generalized coherent states; middle row)
and a
high-energy mess (bottom row).
\label{fftFig} 
}
\end{figure}

A common objection is that repeated branching would exponentially increase the number of
universes over time. However, the number of universes $N$ may well stay constant.
By the number of ``universes'' $N$, we mean the number that are indistinguishable from 
the frog perspective  (from the bird perspective, there is of course just one) at a given instant, \ie, the number 
of macroscopically different Hubble volumes. Although there is obviously a vast 
number of them (imagine moving planets to random new locations, imagine 
having married someone else, \etc), the number $N$ is clearly finite --- even if we pedantically 
distinguish Hubble volumes at the quantum level to be overly conservative, 
there are ``only'' about $10^{10^{115}}$ with temperature 
below $10^8$K as detailed above.\footnote{
For the technical reader, could the grand superposition of the universal wavefunctional
involve other interesting states besides the semiclassical ones?
Specifically, the semiclassical states (corresponding to what we termed the
consensus view) are those 
that are maximally robust towards decoherence (Zurek 2003), so if we project out
the component of the wavefunctional that is spanned by these states, 
what remains?
We can make a hand-waving argument that all that remains is a rather uninteresting
high-energy mess which will be devoid of observers and rapidly expand or collapse.
Let us consider the special case of the electromagnetic field.
In many circumstances (Anglin \& Zurek 1996), its semiclassical states can be 
shown to be generalized coherent states, which have infinite-dimensional Gaussian Wigner
functions with characteristic widths no narrower than the those corresponding to the 
local temperature. Such functions form a well-conditioned basis for 
all states whose wavefunction is correspondingly smooth, \ie, lacking 
violent high-energy fluctuations.
This is illustrated in \fig{fftFig} for the simple case of a 1-dimensional 
quantum particle: the wavefunction $\psi(x)$ can be written as a superposition
of a low energy (low-pass filtered) and a high-energy (high-pass filtered) 
part, and the former can be decomposed as the 
convolution of a smooth function with a Gaussian, \ie, as a superposition
of coherent states with Gaussian wavepackets.
Decoherence rapidly makes the macroscopically distinct semiclassical 
states of the electromagnetic field for all 
practical purposes separate both from each other and from the high-energy mess.
The high-energy component may well be typical of the early universe that we evolved from.
}
The smooth unitary evolution of the wavefunction in the bird perspective
corresponds to a never-ending sliding between 
these $N$ classical universe snapshots from the frog perspective of an observer.
Now you're in universe A, the one where you're reading this sentence.
Now you're in universe B, the one where you're reading this other sentence.
Put differently, universe B has an observer identical to one in universe A,
except with an extra instant of memories.
In \fig{CartoonFig}, our observer first finds herself in the universe described by the left panel,
but now there are two different universes smoothly connecting to it like B did to A, and
in both of these, she will be unaware of the other one.
Imagine drawing a separate dot corresponding to each possible universe and drawing arrows
indicating which ones connect to which in the frog perspective.
A dot could lead uniquely to one other dot or to several, as above.
Likewise, several dots could lead to one and the same dot, since there could be many different
ways in which certain situations could have come about. 
The Level III multiverse thus
involves not only splitting branches but merging branches as well.

Ergodicity implies that the quantum state of the Level III multiverse is invariant
under spatial translations, which is a unitary operation just as time translation.
If it is invariant under time-translation as well (this can be arranged by constructing
a superposition of an infinite set of quantum states that are all 
different time translations of one and the same state, so that a Big Bang happens at different times in different quantum branches), 
then the number of universes would automatically stay exactly constant.
All possible universe snapshots would exist at every instant, 
and the passage of time would just be in the eye of the beholder 
--- an idea explored in the sci-fi novel ``Permutation City'' (Egan 1995)
and developed by Deutsch (1997), Barbour (2001) and others.

\begin{figure}[pbt]
\hglue-1.7cm\centerline{{\vbox{\epsfxsize=12.0cm\epsfbox{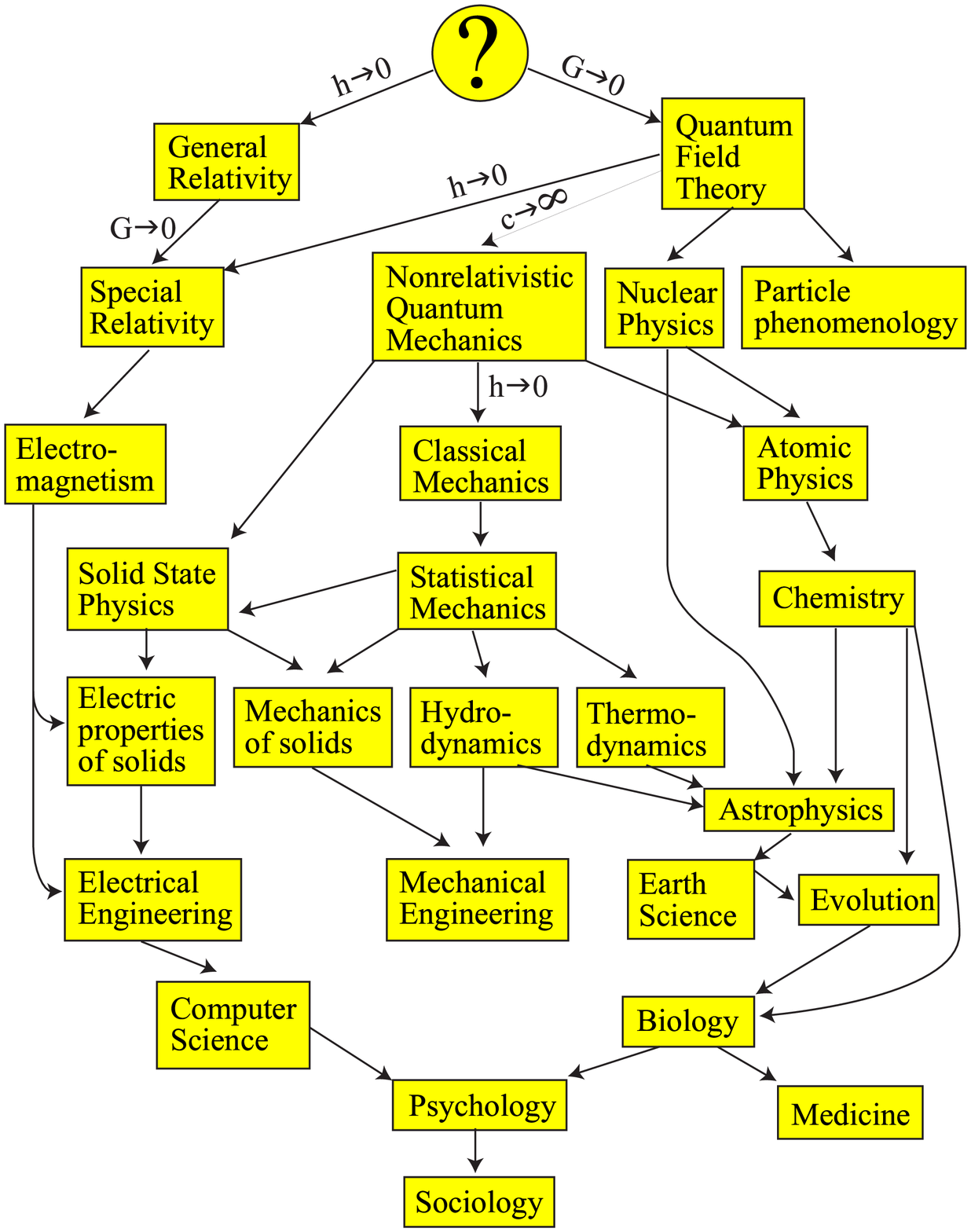}}}}
\smallskip
\caption{
Theories can be crudely organized into a family tree where
each might, at least in principle, be derivable
from more fundamental ones above it.
For example, classical mechanics can be obtained from 
special relativity in the approximation that the speed of light 
$c$ is infinite. Most of the arrows are less well understood.
All these theories have two components: mathematical equations and words
that explain how they are connected to what we observe. 
At each level in the hierarchy of theories, new words (e.g.,
protons, atoms, cells, organisms, cultures) are introduced because they
are convenient, capturing the essence of what is going on without
recourse to the more fundamental theory above it.  It is important to
remember, however, that it is we humans who introduce these concepts and
the words for them: in principle, everything could have been derived from
the fundamental theory at the top of the tree, although such an extreme
reductionist approach would of course be useless in practice. Crudely
speaking, the ratio of equations to words decreases as we move down the
tree, dropping near zero for highly applied fields such as medicine and
sociology. In contrast, theories near the top are highly mathematical,
and physicists are still struggling to understand the concepts, if any,
in terms of which we can understand them.
The Holy Grail of physics is to find what is jocularly referred to as a
``Theory of Everything'', or TOE, from which all else can be derived. If
such a theory exists at all, it should replace the big question mark at
the top of the theory tree. Everybody knows that something is missing
here, since we lack a consistent theory unifying gravity with quantum
mechanics. 
}
\label{TheoryTreeFig}
\end{figure}

\subsection{Two world views}

The debate over how classical mechanics emerges from quantum mechanics
continues, and the decoherence discovery has shown that there is 
a lot more to it than just letting Planck's constant $\hbar$ shrink to zero.
Yet as \fig{TheoryTreeFig}  illustrates, this is just a small piece of a larger puzzle.
Indeed, the endless debate over the interpretation of quantum mechanics 
--- and even the broader issue of parallel universes --- is in a sense the
tip of an iceberg.
In the Sci-Fi spoof ``Hitchhiker's Guide to the Galaxy'', the answer 
is discovered to be ``42'', and the hard part is finding the real question.
Questions about parallel universes may seem to be just about as deep as queries about 
reality can get. Yet there is a still deeper underlying question:
there are two tenable but diametrically opposed
paradigms regarding physical reality and the status of mathematics, a dichotomy that  
arguably goes as far back as Plato and Aristotle, and the question is which one is correct.
\begin{itemize}
\item {\bf ARISTOTELIAN PARADIGM:} The subjectively perceived frog perspective is physically real, 
and the bird perspective and all its mathematical language is 
merely a useful approximation.
\item {\bf PLATONIC PARADIGM:} The bird perspective (the mathematical structure) is physically 
real, and the frog perspective and all the human language we use
to describe it is merely a useful approximation for describing
our subjective perceptions.
\end{itemize}
What is more basic --- the frog perspective or the bird perspective?
What is more basic --- human language or mathematical language?
Your answer will determine how you feel about parallel universes.
If you prefer the Platonic paradigm, you should find multiverses natural, 
since our feeling that say the Level III multiverse is ``weird'' merely reflects that
the frog and bird perspectives are extremely different. We break the symmetry by calling the 
latter weird because we were all indoctrinated with the Aristotelian 
paradigm as children, long before we even heard of mathematics - the Platonic view is an acquired taste!

In the second (Platonic) case, all of physics is ultimately a mathematics problem, since 
an infinitely intelligent mathematician given the fundamental equations of
the cosmos could in principle
{\it compute} the frog perspective, {\ie}, 
compute what self-aware observers the universe would contain, 
what they would perceive, and what language they would 
invent to describe their perceptions to one another. 
In other words, there is a  ``Theory of Everything" (TOE) at the top of the tree in \fig{TheoryTreeFig} 
whose axioms are purely mathematical, since postulates in English 
regarding interpretation would be derivable and thus redundant.
In the Aristotelian paradigm, on the other hand, there can never be a 
TOE, since one is ultimately just explaining
certain verbal statements by other verbal statements ---
this is known as the infinite regress problem (Nozick 1981).

\section{Level IV: Other mathematical structures}

Suppose you buy the Platonist paradigm and believe that
there really is a TOE at the top of \fig{TheoryTreeFig}  --- and that we simply have not found
the correct equations yet.
Then an embarrassing question remains, as emphasized by John Archibald
Wheeler: {\it  Why these particular equations, not others?}
Let us now explore the idea of mathematical democracy, whereby
universes governed by other equations are equally real.
This is the Level IV multiverse.
First we need to digest two other ideas, however:
the concept of a mathematical structure, and the notion that the physical world may be one.

\subsection{What is a mathematical structure?}

Many of us think of mathematics as a bag of tricks 
that we learned in school for manipulating numbers.
Yet most 
mathematicians have a very different view of their field.
They study more abstract objects such as functions, sets, spaces and operators
and try to prove theorems about the relations between them.
Indeed, some modern mathematics papers are so abstract that the only numbers you will
find in them are the page numbers!
What does a dodecahedron have in common with a set of complex numbers?
Despite the plethora of mathematical structures with intimidating names like 
orbifolds and Killing fields, a striking underlying unity that has emerged in the last century:
{\it all} mathematical structures are just special cases of one and the same 
thing: so-called formal systems.
A formal system consists of abstract symbols and rules for manipulating them, 
specifying how new strings of symbols referred to as theorems can be derived from
given ones referred to as axioms.
This historical development represented a form of deconstructionism, since it
stripped away all meaning and interpretation that had traditionally been given to
mathematical structures and distilled out only the 
abstract relations capturing their very essence.
As a result, computers can now prove theorems about geometry
without having any physical intuition whatsoever about what space is like.

\begin{figure}[pbt]
\centerline{{\vbox{\epsfxsize=8.5cm\epsfbox{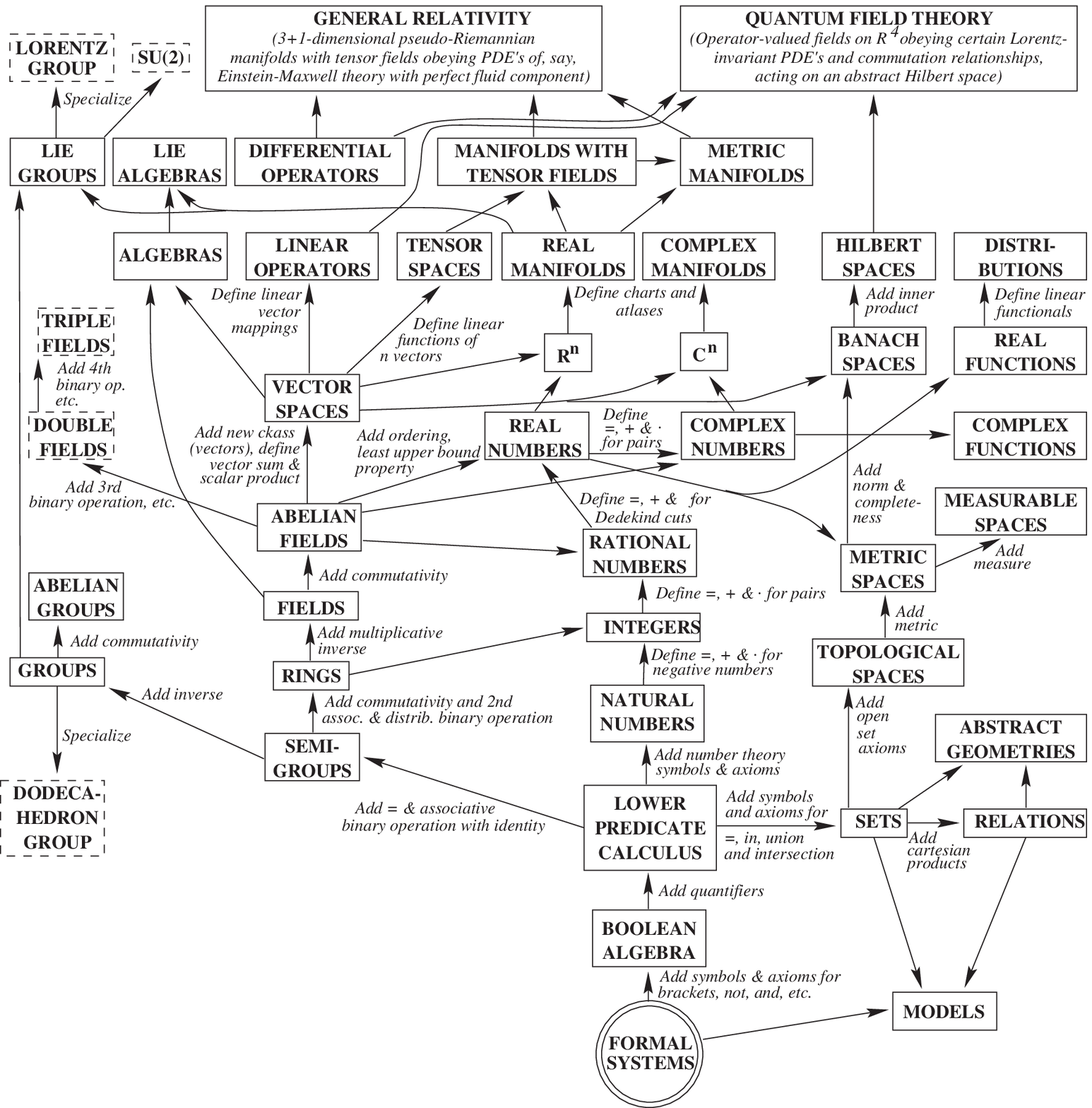}}}}
\smallskip
\caption{
\label{MathTreeFig}
Relationships between various basic mathematical structures (Tegmark 1998).
The arrows generally indicate addition of new symbols and/or axioms.
Arrows that meet indicate the combination of structures
--- for instance,
an algebra is a vector space that is also a ring,
and a Lie group is a group that is also a manifold.
The full tree is probably infinite in extent --- the figure shows
merely a small sample near the bottom.
}
\end{figure}

\Fig{MathTreeFig} shows some of the most basic mathematical structures and
their interrelations. Although this family tree probably extends indefinitely, it illustrates
that there is nothing fuzzy about mathematical structures. 
They are ``out there'' in the sense that mathematicians discover them rather than create them,
and that contemplative alien civilizations would find the same structures
(a theorem is true regardless of whether it is proven by a human, a computer or an alien).

\subsection{The possibility that the physical world is a mathematical structure}

Let us now digest the idea that physical world (specifically, the Level III multiverse)  
{\it is} a mathematical structure.
Although traditionally taken for granted by many theoretical physicists, this is a deep 
and far-reaching notion. It means that mathematical equations describe not merely some
limited aspects of the physical world, but {\it all} aspects of it.
It means that there is some mathematical structure that is what mathematicians call
{\it isomorphic} (and hence equivalent) to our physical world, with each physical
entity having a unique counterpart in the mathematical structure and vice versa.
Let us consider some examples.

A century ago, when classical physics still reigned supreme, 
many scientists believed that physical space was isomorphic 
to the mathematical structure known as $\R^3$:
three-dimensional Euclidean space. 
Moreover, some thought that all forms of matter in the universe
corresponded to various classical {\it fields}: the electric field, the magnetic field and 
perhaps a few undiscovered ones, mathematically corresponding to functions on $\R^3$
(a handful of numbers at each point in space).
In this view (later proven incorrect), dense clumps of matter like atoms were simply 
regions in space where some fields were strong (where some numbers were large).
These fields evolved deterministically over time according to some partial differential equations,
and observers perceived this as things moving around and events taking place.
Could, then, fields in three-dimensional space be the mathematical structure corresponding
to the universe? No,  since a mathematical structure cannot change --- it is an abstract,
immutable entity existing outside of space and time. 
Our familiar frog perspective of a three-dimensional space where events unfold
is equivalent, from the bird perspective, to a four-dimensional spacetime where all of
history is contained, 
so the mathematical 
structure would be fields in four-dimensional space.
In other words, if history were a movie, the mathematical structure would 
not correspond to a single frame of it, but to the entire videotape.

Given a mathematical structure, we will say that it has {\it physical existence}
if any self-aware substructure (SAS) within it subjectively, from its frog perspective,
perceives itself as living in a physically real world.
What would, mathematically, such an SAS be like?
In the classical physics example above, an SAS such as you would be 
a tube through spacetime, a thick version of what Einstein referred to as a  world-line.
The location of the tube would specify your position in space at different times.
Within the tube, the fields would exhibit certain complex behavior, corresponding
to storing and processing information about the field-values in the surroundings,
and at each position along the tube, these processes would give rise to 
the familiar but mysterious sensation of self-awareness.
From its frog perspective, the SAS would perceive this one-dimensional 
string of perceptions along the tube as passage of time.

Although our example illustrates the idea of how our physical world can {\it be} a mathematical
structure, this particular mathematical structure (fields in four-dimensional space) is now known
to be the wrong one. After realizing that spacetime could be curved, Einstein doggedly searched
for a so-called unified field theory where the universe was what mathematicians call 
a 3+1-dimensional pseudo-Riemannian manifold with tensor fields, but this failed to
account for the observed behavior of atoms. According to quantum field theory, the
modern synthesis of special relativity theory and quantum theory, 
the universe (in this case the Level III multiverse) 
is a mathematical structure known as an algebra of operator-valued 
fields. Here the question of what constitutes an
SAS is more subtle (Tegmark 2000).
However, this fails to describe black hole evaporation, the first instance of the Big Bang 
and other quantum gravity phenomena, so 
the true mathematical structure isomorphic to our universe, if it exists, has not yet been found.

\subsection{Mathematical democracy}

Now suppose that our physical world really is a mathematical structure, 
and that you are an SAS within it. This means that in
the Mathematics tree of \fig{MathTreeFig}, one of the boxes is our universe.
(The full tree is probably infinite in extent, so our particular box is not one of the few boxes 
from the bottom of the tree that are shown.)
In other words, this particular mathematical structure enjoys not only mathematical
existence, but physical existence as well.
What about all the other boxes in the tree? Do they too enjoy physical existence? If not, there would be a fundamental, unexplained ontological asymmetry built into the
very heart of reality, splitting mathematical structures into two classes: 
those with and without physical existence. 
As a way out of this philosophical conundrum, I have suggested (Tegmark 1998)
that complete mathematical democracy holds: that mathematical existence and 
physical existence are equivalent, so that  {\it all} mathematical structures exist 
physically as well.
This is the Level IV multiverse. It can be 
viewed as a form of radical Platonism,
asserting that the mathematical structures
in Plato's {\it realm of ideas}, the {\it Mindscape} of Rucker (1982), exist 
``out there'' in a physical sense (Davies 1993),
casting the so-called modal realism theory of  
David Lewis (1986) in mathematical terms
akin to what Barrow (1991; 1992) refers to as ``$\pi$ in the sky''.
If this theory is correct, then since it has no free parameters,
all properties of all parallel universes
(including the subjective perceptions of SASs in them) could in principle be derived by
an infinitely intelligent mathematician.

\subsection{Evidence for a Level IV multiverse}

We have described the four levels of parallel universes in order of increasing
speculativeness, so why should we believe
in Level IV? Logically, it rests on two separate assumptions:
\begin{itemize}
\item {\bf Assumption 1:} That the physical world (specifically our level III multiverse) is a mathematical structure
\item {\bf Assumption 2:} Mathematical democracy: that all mathematical structures  exist ``out there'' in the same sense
\end{itemize}

In a famous essay, Wigner (1967) argued that 
``the enormous usefulness of mathematics in the
natural sciences is something bordering on the mysterious", and that
``there is no rational explanation for it".
This argument can be taken as support for assumption 1: here
the utility of mathematics for describing the physical
world is a natural consequence of the fact that the latter {\it is}
a mathematical structure, and we are simply uncovering this 
bit by bit.
The various approximations that constitute our current physics theories
are successful because simple mathematical
structures can provide good approximations
of how a SAS will perceive more complex mathematical structures.
In other words, our successful theories are
not mathematics approximating physics,
but mathematics approximating mathematics.
Wigner's observation is unlikely to be based on fluke coincidences,
since far more mathematical regularity in nature has been discovered in the 
decades since he made it, including the standard model of particle physics.

A second argument supporting assumption 1 is 
that abstract mathematics is so general that 
{\it any} TOE that is definable in purely formal terms
(independent of vague human terminology)
is also a mathematical structure.
For instance,
a TOE involving a set of different types of entities
(denoted by words, say) and
relations between them (denoted by additional words)
is nothing but what mathematicians call
a set-theoretical model, and one can generally find a
formal system that it is a model of.

This argument also makes assumption 2 more appealing,
since it implies that {\it any} conceivable parallel universe theory can be described 
at Level IV. The Level IV multiverse, termed the ``ultimate Ensemble theory'' in 
Tegmark (1997) since it subsumes
all other ensembles, therefore brings closure to the 
hierarchy of multiverses, and there cannot be say a Level V.
Considering an ensemble of mathematical structures does not add anything new, since
this is still just another mathematical structure.
What about the frequently discussed notion that the universe is a computer simulation?
This idea occurs frequently in science fiction and has been substantially
elaborated (\eg, Schmidthuber 1997; Wolfram 2002).
The information content (memory state) of a digital computer is a string of bits, say 
``$1001011100111001...$'' of great but finite length, equivalent to some large 
but finite integer $n$ written in binary. The information processing of a computer is
a deterministic rule for changing each memory state into another (applied over and over again),
so mathematically, it is simply a function $f$ mapping the integers onto themselves
that gets iterated: $n\mapsto f(n)\mapsto f(f(n))\mapsto...$.
In other words, even the most sophisticated computer simulation is
just yet another special case of a mathematical structure,
and is already included in the Level IV multiverse.
(Incidentally, iterating continuous functions rather than integer-valued ones
can give rise to fractals.)

Another appealing feature of assumption 2 is that it
provides the only answer so far to 
Wheeler's question:
{\it  Why these particular equations, not others?}
Having universes dance to the tune of all possible equations
also resolves the fine-tuning problem of \sec{TuningSec} once and for all,
even at the fundamental equation level:
although many if not most mathematical structures are likely to be dead and devoid
of SASs, failing to provide the complexity, stability and predictability that SASs require,
we of course expect to find with 100\% probability that we inhabit a
mathematical structure capable of supporting life.
Because of this selection effect, the answer to the question
``what is it that breathes fire into the equations and
makes a universe for them to describe?'' (Hawking 1993)
would then be ``you, the SAS''.

\subsection{What are Level IV parallel universes like?}

The way we use, test and potentially rule out any theory is to
compute probability distributions for our future perceptions
given our past perceptions and to compare these predictions with our observed outcome.
In a multiverse theory, there is typically more than one SAS that 
has experienced a past life identical to yours, so there is no way to determine 
which one is you. To make predictions, you therefore have to compute 
what fractions of them 
will perceive what in the future, which
leads to the following predictions:
\begin{itemize}
\item {\bf Prediction 1:}
The mathematical structure describing our world
is the most generic one that is consistent
with our observations.
\item {\bf Prediction 2:}
Our future observations are the most generic ones that
are consistent with our past observations.
\item {\bf Prediction 3:}
Our past observations are the most generic ones that
are consistent with our existence.
\end{itemize}
We will return to the problem of what ``generic'' means in \\sec{MeasureSec} (the measure problem).
However, one 
striking feature of mathematical structures,
discussed in detail in Tegmark (1997), is that the sort of symmetry and invariance properties
that are responsible for the simplicity and orderliness of our universe tend to be
generic, more the rule than the exception --- mathematical structures 
tend to have them by default, and complicated additional axioms {\etc} 
must be added to make them go away. In other words, because of both this and selection effects, 
we should not necessarily expect life in the Level IV multiverse to be a disordered mess.

\section{Discussion}

We have surveyed scientific theories of parallel universes, and found that they naturally
form a four-level hierarchy of multiverses (Figure 1) allowing progressively greater differences
from our own universe:
\begin{itemize}
\item Level I: Other Hubble volumes have different initial conditions
\item Level II: Other post-inflation bubbles may have different 
effective laws of physics (constants, dimensionality, particle content)
\item Level III: Other branches of the quantum wavefunction add nothing qualitatively new
\item Level IV: Other mathematical structures have different fundamental equations of physics 
\end{itemize}
Whereas the Level I universes join seemlessly, there are clear demarcations
between those within levels II and III caused by 
inflating space and decoherence, respectively. The level IV universes are completely 
separate and need to be considered together only 
for predicting your future, since ``you'' may exist in more than one of them.

Although it was Level I that got Giordano Bruno in trouble with the inquisition,
few astronomers today would suggest that space ends 
abruptly at the edge of the observable universe.
It is ironic and perhaps due to historic coincidence
that Level III is the one that has drawn the most fire in the past decades, since it
is the only one that adds no qualitatively new types of universes.

\subsection{Future prospects}

There are ample future prospects for testing and perhaps ruling out these multiverse theories.
In the coming decade, dramatically improved cosmological measurements of the 
microwave background radiation, the large-scale matter distribution, \etc, will test Level I 
by further constraining the curvature and topology of space and will test level II by providing 
stringent tests of inflation.
Progress in both astrophysics and high-energy physics should also clarify the extent to which various
physical constants are fine-tuned, thereby weakening or strengthening the case for Level II.
If the current world-wide effort to build quantum computers succeeds, it will provide 
further evidence for Level III, since they would, in essence, be 
exploiting the parallelism of the Level III multiverse for parallel 
computation (Deutsch 1997). 
Conversely, experimental evidence of unitarity violation would rule out Level III.
Finally, success or failure in the grand challenge of modern physics,  
unifying general relativity and quantum field theory, will shed more light on Level IV. 
Either we will eventually find a mathematical structure matching our universe, or we will 
bump up against a limit to  the unreasonable effectiveness of mathematics 
and have to abandon Level IV.

\subsection{The measure problem}
\label{MeasureSec}

There are also interesting theoretical issues to resolve within the multiverse theories,
first and foremost the {\it measure problem}.
As multiverse theories gain credence,
the sticky issue of how to compute probabilities in physics is growing 
from a minor nuisance into a major embarrassment.
The reason why probabilities become so important is that 
if there are indeed many copies of ``you'' with identical past lives and memories, 
you could not compute your own future even if you had complete knowledge
of the entire state of the multiverse. This is because there is no way for you to determine 
which of these copies is ``you'' (they all feel that they are).
All you can predict is therefore probabilities for what you will observe,
corresponding to the fractions of these observers that experience different things.
Unfortunately, computing what fraction of the infinitely many observers perceive what is very subtle, 
since the answer depends on the order in which you count them!
The fraction of the integers that are even is 50\% if you 
order them 1, 2, 3, 4..., but approaches 100\% 
if you order them alphabetically the 
way your word processor would (1, 10, 100, 1000, ...).
When observers reside in disconnected universes, there is no obviously natural way
in which to order them, and one must sample from the different universes with some
statistical weights referred to by mathematicians as a ``measure''.
This problem crops up in a mild and treatable manner in Level I, becomes severe at Level II,
has caused much debate within the context of extracting quantum probabilities in Level III
(de Witt 2003), 
and is horrendous at Level IV.
At Level II, for instance, Vilenkin and others have published predictions for the probability distributions of
various cosmological parameters by arguing that different parallel universes that have 
inflated by different amounts should be
given statistical weights proportional to their volume (\eg, Garriga \& Vilenkin 2001a).
On the other hand, any mathematician will tell you that $2\times\infty=\infty$, so that
there is no objective sense in which an infinite universe that that has expanded by a factor
of two has gotten larger. 
Indeed, an exponentially inflating universe has what mathematicians call
a time-like Killing vector, which means that it is time-translationally invariant and hence unchanging
from a mathematical point of view.
Moreover, a flat universe with finite volume and the topology of a torus is equivalent to
a perfectly periodic universe with infinite volume, both from the mathematical bird perspective
and from the frog perspective of an observer within it, so why should its infinitely smaller volume
give it zero statistical weight? Since Hubble volumes start repeating even in the Level I multiverse 
(albeit in a random order, not periodically) after about $10^{10^{115}}$ meters, should 
infinite space really be given more statistical weight than a finite region of that size?
This problem must be solved to observationally test models of stochastic inflation.
If you thought that was bad, consider the problem of assigning statistical weights
to different mathematical structures at Level IV. The fact that our universe
seems relatively simple has led many people to suggest 
that the correct measure somehow involves complexity. For instance, one could reward simplicity
by weighting each mathematical structure by $2^{-n}$, where $n$ is its 
algorithmic information content measured in bits, defined as the length of the shortest bit string
(computer program, say) that would specify it (Chaitin 1987). 
This would correspond to equal weights for all infinite bit 
strings (each representable as a real number like $.101011101...$), not for all mathematical structures.
If there is such an exponential penalty for high complexity, we should probably expect to find ourselves
inhabiting one of the simplest mathematical structures complex enough to contain observers.
However, the algorithmic complexity depends on 
how structures are mapped to bit strings (Chaitin 1987; Deutsch 2003), 
and it far from obvious whether there exists a most natural 
definition that reality might subscribe to.

\subsection{The pros and cons of parallel universes}

So should you believe in parallel universes?
Let us conclude with a brief discussion of arguments pro and con.
First of all, we have seen that this is not a yes/no question --- rather, 
the most interesting issue is
whether there are 0, 1, 2, 3 or 4 levels of multiverses.
Figure 1 summarizes evidence for the different levels.
Cosmology observations support Level I by pointing 
to a flat infinite space with ergodic matter distribution,
and Level I plus inflation elegantly eliminates the initial condition problem.
Level II is supported by the success of inflation theory in explaining cosmological observations,
and it can explain apparent fine-tuning of physical parameters.
Level III is supported by both experimental and theoretical evidence for unitarity, and
explains the apparent quantum randomness that bothered Einstein so much without
abandoning causality from the bird perspective. 
Level IV explains Wigner's unreasonable effectiveness of mathematics for describing physics 
and answers the question ``why these equations, not others?".

The principal arguments against parallel universes are that they are wasteful and weird,
so let us consider these two objections in turn. 
The first argument is that multiverse theories are
vulnerable to Ockham's razor,
since they postulate the existence of other worlds that
we can never observe.
Why should nature be so ontologically wasteful and indulge in such opulence
as to contain an infinity of different worlds?
Intriguingly, this argument can be turned around to argue
{\it for} a multiverse. When we feel that nature is wasteful, 
what precisely
are we disturbed about her wasting? Certainly not ``space",
since the standard flat universe model with its infinite volume draws no such objections.
Certainly not ``mass" or ``atoms" either, for the same reason --- once you have
wasted an infinite amount of something, who cares if you waste some more?
Rather, it is probably the apparent reduction in simplicity that
appears disturbing, the quantity of information necessary to specify
all these unseen worlds.
However, as is discussed in more detail in Tegmark (1996),
an entire ensemble is often much simpler than one of its members.
For instance, the algorithmic information content of a
generic integer $n$ is of order $\log_2 n$ (Chaitin 1987), the number of bits required
to write it out in binary.
Nonetheless, the set of all integers $1, 2, 3, ...$
can be generated by quite a
trivial computer program, so the algorithmic
complexity of the whole set is smaller than that of a generic member.
Similarly, the set of all perfect fluid solutions to the
Einstein field equations has a smaller algorithmic complexity than
a generic particular solution, since the former is specified simply by
giving a few equations and the latter requires the specification of
vast amounts of initial data on some hypersurface.
Loosely speaking, the apparent information content rises when
we restrict our attention to one particular element in an ensemble,
thus losing the symmetry and simplicity that was inherent in the totality
of all elements taken together.
In this sense, the higher level multiverses have less algorithmic complexity.
Going from our universe to the Level I multiverse eliminates the need to specify
initial conditions, upgrading to Level II eliminates the need to specify physical 
constants and the Level IV multiverse  
of all mathematical structures
has essentially no algorithmic complexity at all.
Since it is merely in the frog perspective, in the subjective perceptions
of observers, that this opulence of information and complexity is really there,
a multiverse theory is arguably
more economical than one endowing only a single ensemble element
with physical existence (Tegmark 1996).

The second common complaint about multiverses is that they are weird.
This objection is aesthetic rather than scientific, and as mentioned above, 
really only makes sense in the Aristotelian world view.
In the Platonic paradigm, one might expect observers to complain that 
the correct TOE was weird if the bird perspective was sufficiently different from the frog perspective, 
and there is every indication that this is the case for us.
The perceived weirdness is hardly surprising, since
evolution provided us with intuition only for the everyday physics that had survival 
value for our distant ancestors.
Thanks to clever inventions, we have glimpsed slightly more
than the frog perspective of our normal inside view, 
and sure enough, we have encountered bizarre phenomena whenever
departing from human scales in any way: at high speeds (time slows down),
on small scales (quantum particles can be at several places at once), on large scales (black holes),
at low temperatures (liquid Helium can flow upward), 
at high temperatures (colliding particles can change identity), {\etc}
As a result, physicists have by and large already accepted that the frog and
bird perspectives are very different, 
A prevalent modern view of quantum field theory
is that the standard model is merely an effective theory, a low-energy limit of 
a yet to be discovered theory that is even more removed from our cozy classical 
concepts (involving strings in 10 dimensions, say). 
Many experimentalists are becoming blas\'e about  producing so many ``weird''
(but perfectly repeatable) experimental results,
and simply accept that the world is a weirder place than we thought
it was and get on with their calculations.

We have seen that a common feature of all four multiverse levels is that the simplest and arguably most
elegant theory involves parallel universes by default, and that one needs
to complicate the theory by adding experimentally unsupported processes and ad hoc 
postulates (finite space, wavefunction collapse, ontological asymmetry, \etc) to explain away 
the parallel universes. Our aesthetic judgement therefore comes down to 
what we find more wasteful and inelegant: many worlds or many words.
Perhaps we will gradually get more used to the weird ways of our cosmos, 
and even find its strangeness to be part of its charm.

\bigskip
{\bf Acknowledgements: }
The author wishes to thank Anthony Aguirre, Aaron Classens, 
Angelica de Oliveira-Costa, George Musser, 
David Raub, Martin Rees, Harold Shapiro and Alex Vilenkin
for stimulating discussions.
This work was supported by
NSF grants AST-0071213 \& AST-0134999,
NASA grants NAG5-9194 \& NAG5-11099,
a fellowship from the David and Lucile Packard Foundation
and a Cottrell Scholarship from Research Corporation.

\section{References} 

\rf\nnn Anglin J R\dualand\nnn Zurek W H;1996;Phys. Rev. D;53;7327

\rfbook;Barbour J B;2001;The End of Time;Oxford Univ. Press;Oxford
 
\rfbook\nnn Barrow J D;1991;Theories of Everything;Ballantine;{New York}

\rfbook\nnn Barrow J D;1992;Pi in the Sky;Clarendon;Oxford
 
\rfbook\nnn Barrow J D\dualand\nnn Tipler F J;1986;The Anthropic Cosmological Principle;Clarendon;Oxford

\rf\nnn Brundrit G B;1979;Q. J. Royal Astr. Soc.;20;37

\rn\nnn Bucher M A\dualand\nnn Spergel D N 1999, {\it Sci. Am.} {\bf 1/1999}

\rfproc\nn Carter B;1974;IAU Symposium 63;\nn Longair S;Reidel;Dordrecht

\rfbook\nnn Chaitin G J;1987;Algorithmic Information Theory;Cambridge U. P;Cambridge

\rfbook\nnnn Davies P C W;1982;The Accidental Universe;Cambridge U. P;Cambridge

\rfbook\nn Davies P;1993;The Mind of God;Touchstone;{New York}

\rfbook\nn Davies P;1996;Are We Alone?;Basic Books;{New York}

\rfbook\nn Deutsch D;1997;The Fabric of Reality;Allen Lane;{New York}

\rfproc\nn Deutsch D;2003;Science and Ultimate Reality: From Quantum to Cosmos;
\nnn Barrow J D, \nnnn Davies P C W\multiand\nnn Harper C L;Cambridge Univ. Press;Cambridge

\rfproc\nn {de Witt} B;2003;Science and Ultimate Reality: From Quantum to Cosmos;
\nnn Barrow J D, \nnnn Davies P C W\multiand\nnn Harper C L;Cambridge Univ. Press;Cambridge

\rfbook\nn Egan G;1995;Permutation City;Harper;{New York}

\rf\nn Garriga J\dualand\nn Vilenkin A;2001a;Phys. Rev. D;64;023507

\rf\nn Garriga J\dualand\nn Vilenkin A;2001b;Phys. Rev. D;64;043511.

\rfbook\nn Giulini D, \nn Joos E, \nn Kiefer C, \nn Kupsch J,
\nnn Stamatescu I O\multiand\nnn Zeh H D;1996;Decoherence and the Appearance
of a Classical World in Quantum Theory;Berlin;Springer

\rf\nn Guth A\dualand\nnn Steinhardt P J;1984;Sci. Am.;250;116

\rfbook\nn Hawking S;1993;A Brief History of Time;Touchstone;{New York}

\rfbook\nn Lewis D;1986;On the Plurality of Worlds;Blackwell;Oxford

\rf\nn Linde A;1994;Sci. Am.;271;32

\rfproc\nn Maldacena J;2003;Science and Ultimate Reality: From Quantum to Cosmos;
\nnn Barrow J D, \nnnn Davies P C W\multiand\nnn Harper C L;Cambridge Univ. Press;Cambridge

\rfbook\nn Nozick R;1981;Philosophical Explanations;Harvard Univ. Press;Cambridge

\rfbook\nnn Rees M J;2002;Our Cosmic Habitat;Princeton Univ. Press;Princeton

\rfbook\nn Rucker R;1982;Infinity and the Mind;Birkhauser;Boston

\rn\nn Schmidthuber J 1997, {\it A Computer Scientist's View of
Life, the Universe, and Everything}, in 
{\it Foundations of Computer Science: Potential-Theory-Cognition, 
Lecture Notes in Computer Science}, C. Freksa, ed., 
(Springer: Berlin), http://www.idsia.ch/~juergen/everything/html.html

\rfbook\nn Smolin L;1997;The Life of the Cosmos;Oxford Univ. Press;Oxford

\rf\nnn Steinhardt P J\dualand\nn Turok N;2002;Science;296;1436

\rf\nn Tegmark M;1996;Found. Phys. Lett.;9;25

\rf\nn Tegmark M;1997;Class. Quant. Grav.;14;L69

\rf\nn Tegmark M;1998;Ann. Phys.;270;1

\rf\nn Tegmark M;2000;Phys. Rev. E;61;4194 

\rf\nn Tegmark M;2002;Science;296;1427

\rf\nn Tegmark M\dualand\nnn Wheeler J A;2001;Sci.Am.;2/2001;{68-75} 

\rfbook\nnn Wigner E P;1967;Symmetries and Reflections;MIT Press;Cambridge

\rfbook\nnn Wolfram S;2002;A New Kind of Science;Wolfram Media;{New York}

\rf\nnn Zeh H D;1970;Found. Phys.;1;69

\rfproc\nn Zurek W;2003;Science and Ultimate Reality: From Quantum to Cosmos;
\nnn Barrow J D, \nnnn Davies P C W\multiand\nnn Harper C L;Cambridge Univ. Press;Cambridge

\end{document}